\documentclass[preprint,times]{aastex7}%
\usepackage{amsmath}
\usepackage{graphicx}
\usepackage{indentfirst}
\usepackage{graphicx}
\usepackage{subcaption}
\usepackage{caption}
\usepackage{multirow}

\begin{document}

\title{Persistent Radio Sources and Multi-wavelength Counterparts of Fast Radio Bursts: Magnetar Wind Nebulae and Bow Shocks in Massive Binary Systems}

\author[0000-0002-2171-9861]{Z. Y. Zhao} 
\affil{School of Astronomy and Space Science, Nanjing University, Nanjing 210023, China}
\email{}

\author[0000-0001-6021-5933]{Q. Wu} 
\affil{School of Astronomy and Space Science, Nanjing University, Nanjing 210023, China}
\email{}

\author[0000-0003-4157-7714]{F. Y. Wang}
\affil{School of Astronomy and Space Science, Nanjing University, Nanjing 210023, China}
\affil{Key Laboratory of Modern Astronomy and Astrophysics (Nanjing University), Ministry of Education, Nanjing 210093, China}
\email{fayinwang@nju.edu.cn}

\author[0000-0002-7835-8585]{Z. G. Dai} 
\affil{Department of Astronomy, School of Physical Sciences, University of Science and Technology of China, Hefei 230026, China}
\email{daizg@ustc.edu.cn}

\correspondingauthor{F. Y. Wang}
\email{fayinwang@nju.edu.cn}

\begin{abstract}
Fast radio bursts (FRBs) are millisecond-duration pulses originating from cosmological
distances. Multi-wavelength counterparts associated with FRBs are important for
unveiling their physical origins. Recent observations provide strong evidence that the sources of some active FRBs are residing in massive star binaries.  In this paper, we study the electromagnetic counterparts of FRBs, including persistent radio sources (PRSs) and multi-wavelength emission from magnetar wind nebulae (MWNe) or bow shocks in massive binary systems.
We find that the PRSs with luminosity $10^{38}-10^{39}$ erg s$^{-1}$ can be generated by young MWN. The age of magnetars is a few decades. The observed long-term variation of flux density for PRSs can be explained by the internal magnetic field decay of magnetars. The bow shock radiation can account for the less luminous PRS with the luminosity $\lesssim 10^{36}$ erg s$^{-1}$. Magnetically powered nebulae may yield detectable keV synchrotron self-Compton (SSC) X-rays out to $\lesssim 20$ Mpc, whereas rotation-powered systems mainly produce GeV SSC emission detectable only within $\lesssim 0.1$ Mpc. For active magnetars in massive binaries, enhanced wind interaction can generate synchrotron X-rays detectable within $\lesssim 50$ Mpc and GeV-TeV inverse-Compton emission detectable within $\lesssim 1-0.1$ Mpc.

\end{abstract}

\keywords{Fast radio burst; Magnetar; Binary}

\section{Introduction}
Fast radio bursts (FRBs) are millisecond-duration, coherent radio pulses originating from cosmological distances, with their physical origins remaining undetermined \citep{Lorimer2007,ZB2023,Wang2026}. FRB 20121102A is the first discovered repeating FRB \citep{Spitler2016}. As the number of detected FRBs has rapidly increased, sources are commonly discussed observationally in terms of repeating FRBs and apparently non-repeating FRBs. Large samples, such as the CHIME/FRB catalog, allow comparisons of their observed properties, such as fluence, duration,  and sky distribution\citep{CHIME/FRBCollaboration2021,Chime/FrbCollaboration2026}. Among repeating FRBs, some sources exhibit very high burst rates, such as 122 h$^{-1}$ for FRB 20121102A \citep{Li2021}, 462 h$^{-1}$ for FRB 20211124A \citep{Zhou2022}, and 729 h$^{-1}$ for FRB 20240114A \citep{ZhangJ2025}. 

Long-term monitoring of active repeating FRBs suggests that these sources may be embedded in complex environments \citep{Mckinven2023,Ng2025,Feng2025}. FRB 20121102A is the first discovered FRB associated with a compact persistent radio source (PRS) \citep{Chatterjee2017}. The DM of FRB 20121102A increased before 2020 \citep{Hessels2019,Oostrum2020,Li2021} and then decreased afterward \citep{WangP2025,Snelders2025}. In contrast, its RM rapidly declines since the first measurement \citep{Michilli2018,Hilmarsson2021,WangP2025,Zhang2026}. The PRS associated with FRB 20121102A exhibits a flat spectrum in the frequency range of 400 MHz to 22 GHz \citep{Chatterjee2017,Marcote2017,Resmi2021,Plavin2022,Rhodes2023,Bhardwaj2025,Yang2024}. The flux variability has been reported based on observations between 2016 and 2023 \citep{Yang2024}. However, based on seven years of $L$-band monitoring, \cite{Bhardwaj2025} found no significant long-term luminosity variation and argued that the observed scatter is consistent with refractive interstellar scintillation (RISS). In this work, we focus on the long-term intrinsic evolution of PRSs and do not include the RISS effect.

For FRB 20190520B, a notably high DM attributed to the host galaxy was detected, with DM$_\mathrm{host}\sim 900$ pc cm$^{-3}$ \citep{Niu2022}. Additionally, the DM was observed to decrease at a significant rate of approximately $-11$ pc cm$^{-3}$ yr$^{-1}$ \citep{Wang2025,Niu2026}. FRB 20190520B is the second FRB associated with compact PRSs detected from 1.3 to 10.0 GHz \citep{Niu2022,Zhang2023,Bhandari2023,Balasubramanian2025}. The PRS luminosity is decreasing over the period of $\sim 4$ yr at frequencies of $1.5-5.5$ GHz \citep{Zhang2023,Balasubramanian2025}. 
The RM of FRB 20190520B is approximately $10^4$ rad m$^{-2}$, with evidence of a rapid reversal in sign \citep{Anna-Thomas2023}. The complex RM variations have been observed for FRB 20201124A \citep{Xu2022} and FRB 20220529 \citep{Li2026}. A similar RM variation and reversal have been found in the Galactic gamma-ray binaries, e.g., PSR B1259$-$63/LS 2883 \citep{Johnston1996,Johnston2005}. 

PRSs have also been detected in association with both FRB 20201124A \citep{Bruni2024} and FRB 20240114A \citep{Bruni2025}. The PRS associated with FRB 20201124A exhibits an inverted spectrum (with a positive spectral index $F_\nu \propto \nu^\alpha, \alpha=1$) at higher frequencies of 6-22 GHz \citep{Bruni2024}. The PRS associated with FRB 20240114A has been reported to show an evolving spectrum rather than a stable flat spectrum, with possible flux-density variations over time \citep{Zhang2025}. Further observations are needed to determine whether these variations are intrinsic or partly affected by scintillation. The spectral break around $\sim 1.6$ GHz has been detected for the PRS associated with FRB 20240114A, which may be caused by the synchrotron self-absorption (SSA) \citep{Zhang2025}. From the SSA frequency, the size of the emission region is constrained to $\sim 0.03$ pc \citep{Zhang2025}. Recently, two promising PRS candidates have been reported in the location of FRB 20181030A and
20190417A \citep{Ibik2024}, and the compact PRS associated with FRB 20190417A has been confirmed via the milliarcsecond localization of EVN Observations \citep{Moroianu2026}

Many environmental models have been proposed to explain their diverse observational properties. These models are mainly divided into two categories: those involving young magnetars in massive binary systems \citep{Wang2022,Zhao2023,Rajwade2023,Lan2024,ZH2025,Shan2026} and those involving single magnetars embedded in pulsar/magnetar wind nebulae (PWN/MWN, see \citealt{Murase2016,Margalit2018,Yang2019,Yang2020,Zhao2021,Rahaman2025}) and supernova remnants (SNRs, see \citealt{Yang2017,Piro2018,Zhao2021a,Katz2022,Zhao2024}). The single magnetar model successfully explains the long-term evolution of extreme DM and RM, as well as the radiation of PRS, but it fails to explain the reversal of RM. The binary model can naturally explain the RM reversals due to orbital motion, but it cannot account for the extreme DM because of free-free absorption from the dense stellar wind. 

Recently, \cite{Wang2025} proposed a unified model, the magnetar-massive-star binary embedded in a SNR, which can explain most DM and RM observations. This study offers an explanation for PRS in binary systems using the unified model framework. When the magnetar wind interacts with the companion wind, the bow shock is generated. For a young magnetar, the shock bends back to the star (see Figure \ref{fig:model}b). In this case, the magnetar wind will propagate freely, especially in the direction perpendicular to the orbital plane \citep{Bosch-Ramon2011}. As the free expanding magnetar wind interacts with the outer SN ejecta, it generates a luminous MWN at a distance much greater than the orbit separation.

Electromagnetic counterparts associated with FRBs are important for unveiling their physical origins \citep{Zhang2024}. To date, the only known electromagnetic counterpart of FRBs is the X-ray bursts from Galactic magnetar SGR J1935$+$2154 \citep{Bochenek2020,CHIME/FRBCollaboration2020}. In addition to prompt counterparts from the central engine, persistent multi-wavelength emission may also provide important clues to the environments of FRB sources, especially if some FRBs reside in binary systems \citep{Wang2022,Zhao2023,Rajwade2023,ZH2025}. In such systems, the interaction between the compact-object outflow and the stellar wind of a massive companion can form a bow shock and produce broadband emission. The multi-wavelength emission (radio to TeV $\gamma-$ray) arising from synchrotron radiation and inverse-Compton (IC) scattering, including synchrotron self-Compton (SSC) scattering of the synchrotron photons and external inverse Compton (EIC) scattering of the stellar photons, in the bow shock is found in high-mass $\gamma-$ray binaries (HMGBs), e.g., PSR B1259$-$63 \citep{Johnston2005,Aharonian2005b,Chernyakova2009,Abdo2011}, LS I $+61^{\circ} 303$ \citep{Marti1995,Chernyakova2006,Acciari2009} and LS 5039 \citep{Marti1998,Clark2001,Martocchia2005,Aharonian2005}.

This paper is organized as follows. The overall description of the binary models is given in Section \ref{sec:model}. In Section \ref{sec:nebula}, we present numerical results for the
spectral evolution model of MWN. Section \ref{sec:binary} describes the multi-wavelength radiation from the bow shock. Section \ref{sec:results} presents the fitted spectrum and light curves for the four PRSs, such as FRB 20121102A, FRB 20190520B, FRB 20201124A, and FRB 20240114A. The contributions of the nebula models to both DM and RM are briefly addressed in Section \ref{sec:dmrm}. Main conclusions and predictions that can be tested by subsequent observations are summarized in Section \ref{sec:con}.

\section{Model description}\label{sec:model}
\subsection{Energy Injection}
In the context of MWN models, the compact PRSs are powered by the rotation or magnetic energy injection from the young magnetar. The initial rotational energy of a magnetar with initial spin period $P_{\mathrm{i}}$ is
\begin{equation}    
E_{\mathrm{rot}}=\frac{1}{2} I \Omega_{\mathrm{i}}^2 \approx 2 \times 10^{52} P_{\mathrm{i},-3}^{-2} \mathrm{erg},
\end{equation}
where $I=10^{45} \mathrm{~g} \mathrm{~cm}^2$ is the moment of inertia, and $\Omega_{\mathrm{i}}=2\pi/P_{\mathrm{i}}$ is the initial spin angular frequency. The spin-down luminosity of a magnetar is
\begin{equation}
L_{\mathrm{sd}} \approx7.2 \times 10^{47} \mathrm{erg}\mathrm{s}^{-1}B_{\mathrm{d}, 14}^2 P_{\mathrm{i},-3}^{-4}\left(1+\frac{t}{t_{\mathrm{sd}}}\right)^{-2},
\end{equation}
where $B_{\mathrm{d}}$ is the dipolar magnetic field and $t_{\mathrm{sd}} \approx 3 \times 10^4 \mathrm{s}~P_{\mathrm{i},-3}^2 B_{\mathrm{d}, 14}^{-2} $ is the spin-down timescale.

The magnetar's internal magnetic field may also power the MWN via energetic outflows. The initial internal magnetic energy of a magnetar with an initial internal magnetic field $B_{\mathrm{int}}$ is 
\begin{equation}\label{eq:EBint}
E_{\mathrm{B,i}}=\frac{1}{6} B_{\mathrm{int}}^2 R_{\mathrm{NS}}^3 \approx 2 \times 10^{49} B_{\mathrm{int}, 16}^2 \mathrm{~erg},
\end{equation}
where $R_{\mathrm{NS}}$ is the magnetar radius. If the source is active enough, the magnetic energy release process can be regarded as continuous, which is a good approximation for a young magnetar.
The internal magnetic field decay luminosity of a magnetar is \citep{Colpi2000,Beniamini2025,Rahaman2025}
\begin{equation}
L_{\mathrm{B}}(t)=L_{\mathrm{B}, 0}\left(1+\frac{\alpha_{\mathrm{B}} t}{t_{\mathrm{d}}}\right)^{-\frac{2}{\alpha_{\mathrm{B}}}-1}.
\end{equation}
Following \cite{Rahaman2025}, the decay index $\alpha_{\mathrm{B}}=1$ is taken throughout in our calculation. In this case, the magnetic field decay timescale is $t_{\mathrm{d}} \approx 10^3 t_{0, 3} B_{\mathrm{int}, 16}^{-\alpha_{\mathrm{B}}} ~\mathrm{yr}$, where $t_{0,3}=t_0 /\left(10^3 ~\mathrm{yr}\right)$ is the fiducial decay timescale for $B_{\mathrm{int}}=10^{16}$ G. 

The initial magnetic field luminosity is
\begin{equation}
L_{\mathrm{B}, 0}=\frac{2 E_{\mathrm{B}, 0}}{t_{\mathrm{d}}} \approx 1.3 \times 10^{39} t_{0, 3}^{-1} B_{\mathrm{int}, 16}^{2+\alpha_{\mathrm{B}}} \mathrm{~erg} \mathrm{~s}^{-1}.
\end{equation}
The timescale for the decay of the fiducial magnetic field is highly uncertain. Research focusing on anomalous X-ray pulsars (AXPs) indicates that the magnetic field undergoes substantial decay over a typical period ranging from $t_0\sim10^{2}-10^{4}$ years \citep{Colpi2000}. In the previous study of PRSs, models with shorter magnetic field decay timescales ($t_0\lesssim1$ yr) have also been proposed \citep{Margalit2018}. We therefore treat $t_0$ as a free parameter, and the magnetic field decay timescale $t_\mathrm{d}$ is derived from $t_0$, $B_{\mathrm{int}}$ and $\alpha_{\mathrm{B}}$.

\subsection{The influence of the companion}
If the magnetar is in a binary system, the interaction between the stellar wind from the companion and the magnetar wind will affect the evolution of the magnetar outflows. It is convenient to describe wind interactions in terms of the momentum flux ratio of the magnetar wind and the stellar wind
\begin{equation}
\begin{aligned}
\eta&=g_j\frac{\dot{E_j}/ c}{\dot{M} v_{\mathrm{w}}}\\
&\simeq 1.7\times 10^3~g_j\dot{E}_{j,40}\left(\frac{\dot{M}}{10^{-8} M_\odot ~\mathrm{yr}^{-1}}\right)^{-1} \left(\frac{v_{\mathrm{w}}}{3\times 10^8~\mathrm{cm~s^{-1}}}\right)^{-1},
\end{aligned}
\end{equation}
where $c$ is the velocity of light, $j=\{\mathrm{sd,B}\}$ represents the cases of spin-down or internal magnetic energy powered magnetar wind. $\dot{E}_j=L_{\mathrm{sd}}$ (or $L_{\mathrm{B}}$) is the spin-down (or magnetic field decay) luminosity of the magnetar and $g_j=\Gamma_j \beta_j/(\Gamma_j-1)$. The bulk Lorentz factor of the magnetar outflows is $\Gamma_j=1/\sqrt{1-\beta_{j}^2}$ with the velocity $v_j=\beta_j c$. For the ultrarelativistic electron-positron pair pulsar wind ($\Gamma_{\mathrm{sd}}\gg 1$), we have $g_\mathrm{sd}\sim 1$. However, the magnetar flare ejecta is baryon-loaded with mildly relativistic bulk velocity \citep{Granot2006}. For the typical magnetar flare ejecta velocity $\beta_{\mathrm{B}} \approx 0.3-0.7$ \citep{Gaensler2005,Granot2006}, we have $g_\mathrm{B}\sim 2.4-6.5$. Therefore, for the estimation of momentum flux, the value $g_j\sim 1$ is a very good approximation. The mass-loss rate of the companion star is $\dot{M}\sim10^{-11}-10^{-8} M_{\odot} ~\mathrm{yr}^{-1}$ for B-type stars \citep{Snow1981,Krticka2014} and $\dot{M}\sim10^{-7}-10^{-5} M_{\odot} ~\mathrm{yr}^{-1}$ for O-type stars \citep{Puls1996,Muijres2012}. The typical stellar wind velocity is $v_{\mathrm{w}}\sim 10^8 \mathrm{~cm} \mathrm{~s}^{-1}$ from observations. 

\begin{figure}
    \centering
    \includegraphics[width=\linewidth]{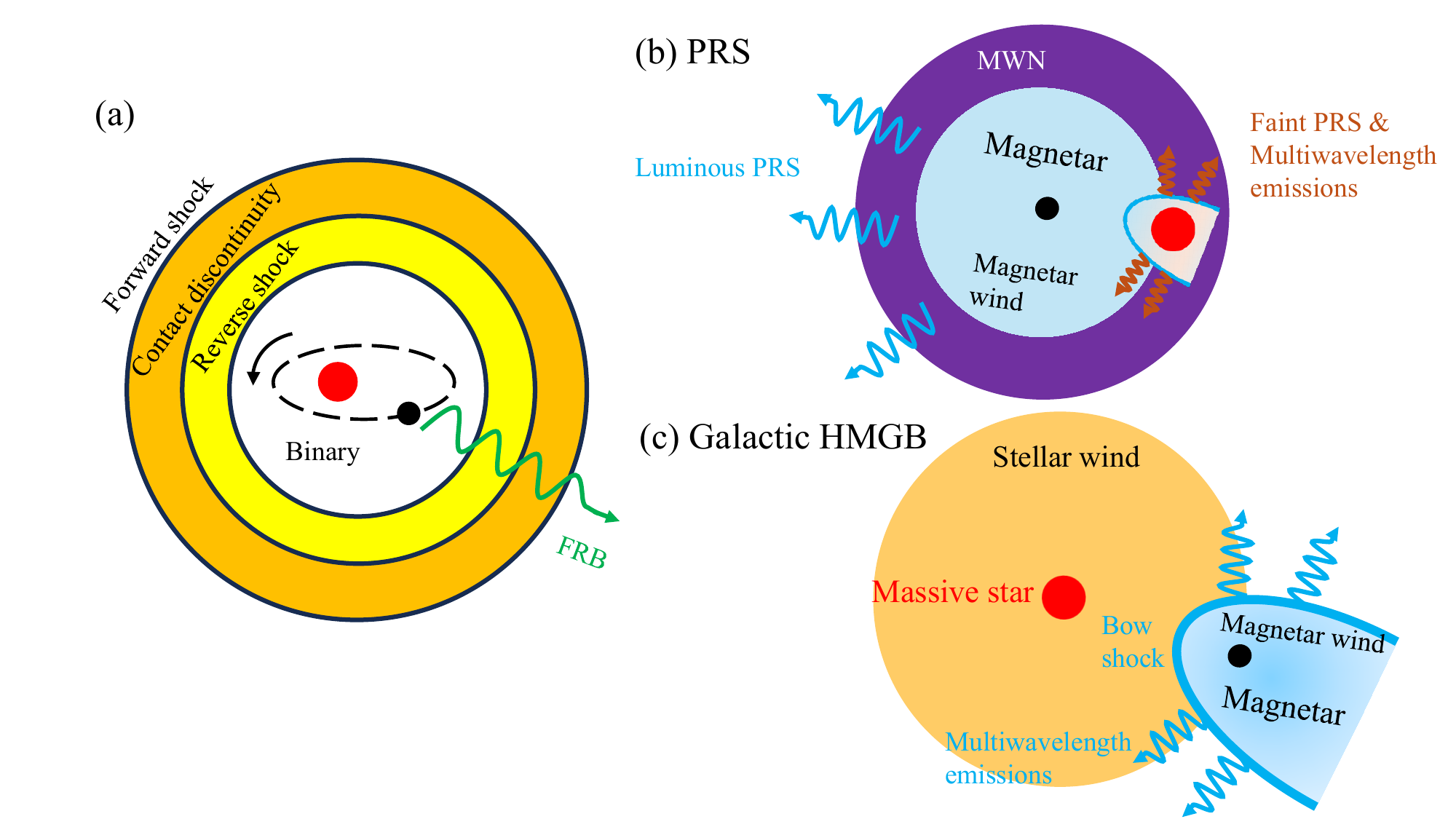} 
    \caption{Schematic diagram of repeating FRBs' environments. \textbf{(a),} The magnetar/massive star binary is embedded in a supernova remnant. \textbf{(b),} The case of a stronger magnetar wind ($\eta \gg 1$), the shock bends back to the star. The magnetar wind will propagate freely, especially in the direction perpendicular to the orbital plane. As the free expanding magnetar wind interacts with the outer SN ejecta, it generates an MWN at a distance much greater than the orbit separation. The observed luminous PRSs are produced by the synchrotron radiation of the MWN. \textbf{(c),} The case of a stronger companion wind ($\eta \ll 1$), the shock bends back to the magnetar. The magnetar wind will be terminated by the stellar wind at the standoff distance $r_{\mathrm{s}}$. The high-energy emission (keV to TeV) arising from synchrotron radiation and inverse-Compton scattering in the bow shock, which can be the multiwavelength counterpart of FRBs. The binary systems are unlikely to be responsible for generating bright PRSs, but they may account for some faint sources, such as FRB 20201124A and FRB 20181030A. Although the magnetar wind is typically stronger than the stellar wind in binary systems containing a young magnetar and a massive star, the wind interaction still occurs within the orbital plane.}
    \label{fig:model}
\end{figure}

When the magnetar wind interacts with the companion wind, the bow shock is generated. The distances from the shock to the magnetar and companion are
\begin{equation}
r_{\mathrm{s}}=d \frac{\eta^{1 / 2}}{1+\eta^{1 / 2}},    
\end{equation}
and 
\begin{equation}
R_{\mathrm{s}}=d \frac{1}{1+\eta^{1 / 2}}.
\end{equation}
The half-opening angle of the bow shock is \citep{Eichler1993}
\begin{equation}\label{eq:theta_h}
\theta_{\mathrm{h}}=2.1\left(1-\bar{\eta}^{2 / 5} / 4\right) \bar{\eta}^{1 / 3},
\end{equation}
where $\bar{\eta}=\min \left(\eta, \eta^{-1}\right)$. For a stronger magnetar wind ($\eta \gg 1$), the shock bends back to the star as shown in Figure \ref{fig:model}b. The half-opening angle of the bow shock is $\theta_{\mathrm{h}}\sim 0.2$ for $\bar{\eta}\sim 0.001$. In this case, the magnetar wind will propagate freely, especially in the direction perpendicular to the orbital plane \citep{Bosch-Ramon2011}. As the free expanding magnetar wind interacts with the outer SN ejecta, it generates a luminous MWN at a distance much greater than the orbit separation. For simplicity, we still assume that the MWN is spherically symmetric and the physical quantities (such as density and magnetic field) in it are uniform. Thus, following the approach outlined in the case of a single magnetar \citep{Murase2016,Margalit2018,Zhao2021,Rahaman2025}, we can derive the model for the long-term evolution of the MWN (see Section \ref{sec:nebula} in detail).

For a stronger companion wind ($\eta \ll 1$), the shock bends back to the magnetar (see Figure \ref{fig:model}c), which is similar to the $\gamma$-ray binary PSR B1259$-$63 \citep{Johnston1996,Johnston2005}. The magnetar wind will be terminated by the stellar wind at the standoff distance $r_{\mathrm{s}}$. The multi-wavelength emission (radio to TeV $\gamma-$ray) arising from synchrotron radiation and IC scattering in the bow shock is found in high-mass $\gamma-$ray binaries (HMGBs), e.g., PSR B1259$-$63 \citep{Johnston2005,Aharonian2005b,Chernyakova2009,Abdo2011}, LS I $+61^{\circ} 303$ \citep{Marti1995,Chernyakova2006,Acciari2009} and LS 5039 \citep{Marti1998,Clark2001,Martocchia2005,Aharonian2005}. The extended radio sources have been discovered from HMGBs, with projected distances up to $100-1000$ AU \citep{Paredes2002,Massi2004,Moldon2011}. The shocked material flows away from the bow shock may produce the pulsar nebula, which can replicate the observed radio morphology \citep{Dubus2006}. However, the radiation of the  bow shock is dominated by high-energy emission, and the radio luminosity is very low \citep{Marti1998,Strickman1998,Johnston2005}. We believe that the binary systems are unlikely to be responsible for generating bright PRSs, but they may account for some faint sources, such as FRB 20201124A \citep{Bruni2024} and FRB 20181030A \citep{Ibik2024}. Although the magnetar wind is typically stronger than the stellar wind in binary systems containing a young magnetar and a massive star, the wind interaction still occurs within the orbital plane. We predict multi-wavelength radiation from the bow shock as the counterpart of FRBs (see Section \ref{sec:binary} in detail).

The bow shock would exist only when $R_{\mathrm{s}}>R_{\star}$ \citep{Wei2024}. The maximum energy-loss luminosity from the rotation-powered engine ($\Gamma_{\mathrm{sd}}\gg 1$) is given by $R_{\mathrm{s}}\sim R_{\star}$, i.e.,
\begin{equation}\label{eq:Lmax} 
\begin{aligned}
L_{\mathrm{max}} & \simeq \dot{M} v_w c\left(\frac{d}{R_\star}-1\right)^2 \\
& \simeq 2.6 \times 10^{39}~\mathrm{erg ~s^{-1}}d_{\mathrm{AU}}^2\left(\frac{\dot{M}}{10^{-8} M_\odot \mathrm{yr}^{-1}}\right) \left(\frac{v_{\mathrm{w}}}{3\times 10^8~\mathrm{cm~s^{-1}}}\right)\left(\frac{R_\star}{10 ~R_{\odot}}\right)^{-2}.
\end{aligned}
\end{equation}
The second line of Equation (\ref{eq:Lmax}) is relevant in the context where $d \gg R_\star$. 

A circular orbit is assumed to estimate the distance, which is given by
\begin{equation}
\begin{aligned}
    d&=a=\left(\frac{G M_{\mathrm{tot}} P_{\mathrm{orb}}^2}{4 \pi^2}\right)^{1 / 3}\\
    & \simeq 1.3~\mathrm{AU} \left(\frac{M_{\mathrm{m}}+M_{\star}}{30~M_\odot}\right)^{1 / 3}\left(\frac{P_{\mathrm{orb}}}{100~\mathrm{d}}\right)^{2 / 3},
\end{aligned}
\end{equation}
where $M_{\mathrm{m}}, M_{\star}$ and $P_{\mathrm{orb}}$ are the mass of the magnetar, the mass of the companion star, and the orbital period, respectively.

The dense stellar wind could absorb FRBs via the free-free absorption. The free-free absorption coefficient at frequency $\nu$ is $\alpha_\nu\simeq0.018 ~\mathrm{cm}^{-1}Z^2 n_{\mathrm{e}} n_{\mathrm{i}} T^{-3 / 2} \nu^{-2} \bar{g}_{\mathrm{ff}}$ \citep{Rybicki1979}, where the Gaunt factor $\bar{g}_{\mathrm{ff}}\sim 1$ and the atomic number of the ion $Z\sim 1$ are used in this work. The electron number density in the stellar wind at the distance $r$ is
\begin{equation}
n_{\mathrm{w}}(r)=n_{\mathrm{w}, 0}\left(\frac{r}{R_{\star}}\right)^{-2},
\end{equation}
where $n_{\mathrm{w}, 0}=\dot{M} / 4 \pi R_{\star}^2 v_{\mathrm{w}} m_{\mathrm{p}}$ is the number density at the stellar surface with $m_{\mathrm{p}}$ being the protons mass. The temperature of the stellar wind depends on adiabatic cooling and photoionization heating \citep{Kochanek1993}. Following the study of radio-absorption in massive binary systems, the adiabatic prescription is adopted \citep{Kochanek1993,Chen2021}
\begin{equation}
T_{\mathrm{w}}(r)=T_{\star}\left(\frac{r}{R_{\star}}\right)^{-2(\hat{\gamma}-1)},
\end{equation}
where $T_{\star}$ is the effective temperature at the surface of the star. The temperature decreases as $T_{\mathrm{w}}\propto r^{-4/3}$ for a gas-pressure-dominated monatomic wind with an adiabatic index $\hat{\gamma}=5/3$.

The free–free optical depth from the stellar wind is 
\begin{equation}
\begin{aligned}
\tau_{\mathrm{ff,w}}&\approx0.018Z^2 \nu^{-2}\left(\frac{\dot{M}}{4 \pi v_\mathrm{w}  \mu_\mathrm{i} m_\mathrm{p}d^2}\right)^2 \cdot T_{\star}^{-3 / 2}\left(\frac{d}{R_{\star}}\right)^{3(\hat{\gamma}-1)} \cdot d\\
&\approx  4.4~d_\mathrm{AU}^{-1}T_{\star,4.5}^{-3 / 2}\nu_{9}^{-2}\left(\frac{\dot{M}}{10^{-8} ~M_\odot \mathrm{yr}^{-1}}\right)^2\left(\frac{v_{\mathrm{w}}}{3\times 10^8~\mathrm{cm~s^{-1}}}\right)^{-2}\left(\frac{R_\star}{10 ~R_{\odot}}\right)^{-1},
\end{aligned}
\end{equation}
for an adiabatic index of $\hat{\gamma}=5/3$. In this work, we take a rotational powered model as an example to estimate the maximum luminosity from the bow shock in a binary system. For the magnetic-powered models, the maximum luminosity is $L_{\mathrm{\max,B}}=L_{\max}\cdot(\Gamma_{\mathrm{B}}-1)/\Gamma_{\mathrm{B}}\beta_{\mathrm{B}}\sim L_{\mathrm{max}}$. The maximum synchrotron luminosity can be estimated as $L_{\mathrm{syn}}= L_{\mathrm{max}}/(1+Y)$.
The luminosity ratio of the IC radiation (including SSC and EIC) to the synchrotron radiation can be described by the Compton parameter $Y$ \citep{Sari2001}. The EIC scattering depends on the thermal photons from the companions. The details of stellar properties and the effect of the IC
process can be found in Appendixes \ref{sec:star} and \ref{sec:ic}.

For PRSs originating from binary systems, the following criteria must be satisfied: (i) the allowed $L_{\mathrm{syn}}$ must exceed PRS luminosity $\nu L_{\nu}$; (ii) the free–free optical depth from stellar wind should be $\tau_{\mathrm{ff,w}}<1$. To determine the parameter space of binary systems that satisfy the above conditions, we employ the empirical scaling relations (see Appendix \ref{sec:star} for details). 
The mass-loss rate of massive stars is taken from the models of \citep{Vink2000}. For solar-like stars, solar mass-loss rate $\dot{M}\sim \dot{M}_{\odot} \approx 2 \times 10^{-14} M_{\odot}~\mathrm{yr}^{-1}$ is used \citep{Wood2002}. 

\begin{figure}
    \centering
    \includegraphics[width=\linewidth]{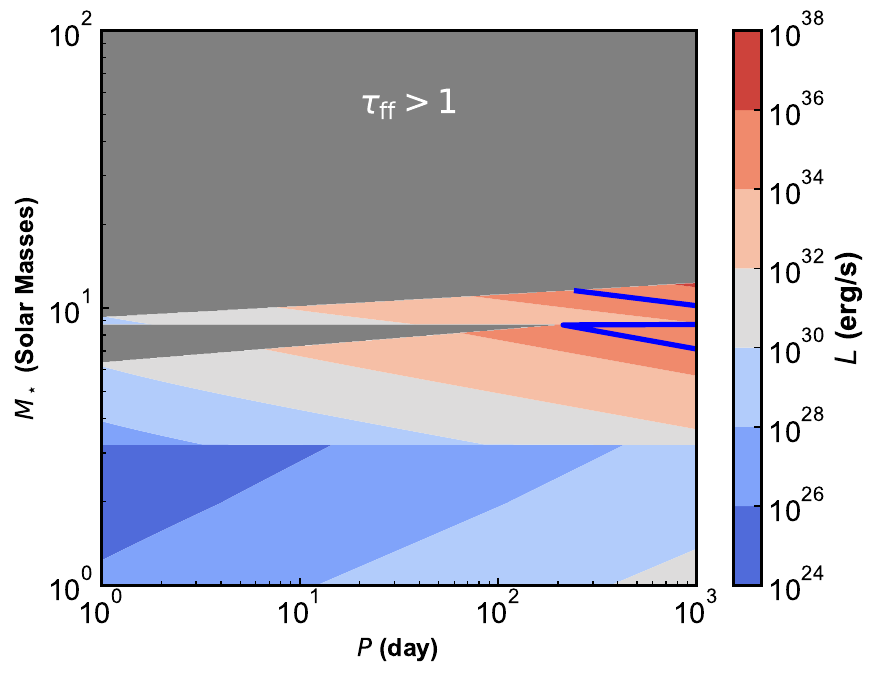}
    \caption{The value of $L_{\mathrm{syn}}$ for different companion mass $M_{\star}$ and orbital periods $P$. The magnetization parameter $\sigma =0.01$ is adopted in our calculation. The free–free absorption optically thick regions for $\nu=1$ GHz have been excluded (shown in gray). The allowed maximum synchrotron luminosity in binary systems is $\sim 10^{36}$ erg s$^{-1}$, which can only account for some of the faintest sources—such as the PRS of FRB 20181030A ($\nu L_{\nu}\sim 10^{35}$ erg s$^{-1}$), shown in blue lines.}
    \label{fig:Lmax}
\end{figure}

The value of $L_{\mathrm{syn}}$ for different companion mass $M_{\star}$ and orbital periods $P$ is shown in Figure \ref{fig:Lmax}. Here, we use the fast-cooling case ($\epsilon_{\mathrm{rad}}=1$) and the Thomson scattering regime ($\epsilon_{\mathrm{KN}}=1$) to estimate the upper limit of the synchrotron radiation. The magnetization parameter $\sigma =0.01$ is adopted in our calculation. The free–free absorption optically thick regions for $\nu=1$ GHz have been excluded (shown in gray). The allowed maximum synchrotron luminosity in binary systems is $\sim 10^{36}$ erg s$^{-1}$, which can only account for some of the faintest sources, such as the PRS of FRB 20181030A ($\nu L_{\nu}\sim 10^{35}$ erg s$^{-1}$), shown in blue lines.

\section{The emission from the rotation/magnetic-powered MWN}\label{sec:nebula}
\subsection{Dynamics and the Nebular Magnetic Field Evolution}
The inner density structure of the unshocked SNe ejecta can be described as a power-law profile \citep{Kasen2010}
\begin{equation}
\rho_{\mathrm{ej}}(r, t)=\frac{(3-\delta)}{4 \pi} \frac{M_{\mathrm{ej}}}{R_{\mathrm{ej}}(t)^3}\left(\frac{r}{R_{\mathrm{ej}}(t)}\right)^{-\delta},   
\end{equation}
where $M_{\mathrm{ej}}$ is the ejecta mass, and $R_{\mathrm{ej}}=v_{\mathrm{ej}}t$ is the radius of the ejecta. The ejesta density profile is usually very flat (with index $\delta=0-1$) due to the mixture of material \citep{Chevalier1977}, and we set $\delta=1$ in this work. In the free expansion (FE) phase, the ejecta expands with a nearly constant velocity of $v_{\mathrm{ej}} = \sqrt{2\mathcal{E}_{\mathrm{SN}}/M_{\mathrm{ej}}}\simeq10^4 \mathrm{~km} \mathrm{~s}^{-1}\left(\mathcal{E}_{\mathrm{SN}} / 10^{51} \mathrm{erg}\right)^{1 / 2}\left(M_{\mathrm{ej}} / M_{\odot}\right)^{-1 / 2}$, where $\mathcal{E}_{\mathrm{SN}}$ is the supernova explosion energy. The dynamical radius of the nebula $R_{\mathrm{n}}$ is given by \citep{Metzger2014}
\begin{equation}\label{eq:dRndt}     
\frac{\mathrm{d} R_\mathrm{n}}{\mathrm{d} t}=\sqrt{\frac{7}{6(3-\delta)} \frac{\mathcal{E}_{\mathrm{inj}}}{M_{\mathrm{ej}}}\left(\frac{R_\mathrm{n}}{R_{\mathrm{ej}}}\right)^{\delta-3}}+\frac{R_\mathrm{n}}{t}
\end{equation}
for $R_{\mathrm{n}}<R_{\mathrm{ej}}$ and $\mathcal{E}_{\mathrm{inj}}=\int \dot{E}_jdt$ is the energy injected into the nebula from the magnetar. Nearly all the rotational energy will be released within a short time $t_{\mathrm{sd}}$ for a magnetar. When the rotational energy exceeds the supernova explosion energy ($\mathcal{E}_{\mathrm{inj}}>\mathcal{E}_{\mathrm{SN}}$), the nebula will be accelerated significantly and the nebula may catch up with the expanding ejecta. In this case, the nebula and ejecta radius will evolve together
\begin{equation}\label{eq:dRndt_f} 
\frac{\mathrm{d} R_{\mathrm{ej}}}{\mathrm{~d} t}=\frac{\mathrm{d} R_{\mathrm{n}}}{\mathrm{~d} t}=v_{\mathrm{ej}, \mathrm{f}},
\end{equation}
where the final velocity is
\begin{equation}
\begin{aligned}
v_{\mathrm{ej,f}}&=\sqrt{2\left(\mathcal{E}_{\mathrm{inj}}+\mathcal{E}_{\mathrm{SN}}\right) / M_{\mathrm{ej}}}  \\
& \approx 0.1 c~ \mathcal{E}_{\mathrm{rot,52}}^{1/2} M_{\mathrm{ej,1M_{\odot}}}^{-1/2}. 
\end{aligned}
\end{equation}

For a millisecond magnetar, the expansion radius of the nebula can be approximated as $R_{\mathrm{n}}=R_{\mathrm{ej}}\simeq v_{\mathrm{ej}, \mathrm{f}}t$ because the acceleration timescale ($\sim t_{\mathrm{sd}}$) is very short. If the energy injection did not cause a significant acceleration of the nebula ($\mathcal{E}_{\mathrm{inj}}<\mathcal{E}_{\mathrm{SN}}$), the expanding velocity is weakly time-dependent \citep{Chevalier1977}. A constant expanding velocity $v_{\mathrm{n}} \lesssim v_{\mathrm{ej}}$ is widely used \citep{Margalit2018,Zhao2021}.



The size of the magnetized emission region is $R_\mathrm{m}=\chi R_{\mathrm{n}}$, which can be smaller than the dynamical radius of the nebula ($\chi \leq 1$, see \citealt{Zhang2008}). The evolution of magnetic energy in the magnetized region depends on the adiabatic expansion and the energy injection from the central engine \citep{Margalit2018}
\begin{equation}\label{eq:Bn}    
\frac{\mathrm{d} \mathcal{E}_{\mathrm{B}}}{\mathrm{d} t}=-\frac{\dot{R}_{\mathrm{m}}}{R_{\mathrm{m}}} \mathcal{E}_{\mathrm{B}}+\frac{\sigma_j}{1+\sigma_j} \dot{E}_{j},
\end{equation}
where $\sigma_j$ is the magnetization parameter of the injected magnetar wind. The magnetic energy in the nebula is $\mathcal{E}_{\mathrm{B}}=B_\mathrm{m}^2 /8\pi \cdot V_\mathrm{m}$, where $B_\mathrm{m}$ and $V_\mathrm{m}=4\pi R_{\mathrm{m}}^3/3$ are the average magnetic field and the volume of the magnetized region. 

Equation (\ref{eq:dRndt}) was introduced as the general dynamical equation for the nebular radius. However, because the time-dependent evolution of the nebular radius (and emitting-region radius) cannot be directly constrained by the available observations, we do not integrate Equation (\ref{eq:dRndt}) in the spectral and light-curve calculations presented in this work. Instead, we adopt a simplified constant expanding velocity ($v_{\mathrm{n}}\simeq v_{\mathrm{ej}, \mathrm{f}}$ for $\mathcal{E}_{\mathrm{inj}}>\mathcal{E}_{\mathrm{SN}}$ and $v_{\mathrm{n}} \lesssim v_{\mathrm{ej}}$ for $\mathcal{E}_{\mathrm{inj}}<\mathcal{E}_{\mathrm{SN}}$) throughout all numerical calculations. The radius of the emitting region must satisfy $R_\mathrm{m}=\chi R_{\mathrm{n}}=\chi v_{\mathrm{n}}t_{\mathrm{age}}< R_{\mathrm{obs}}$, where $t_{\mathrm{age}}$ is the source age and $R_{\mathrm{obs}}$ is the upper limit inferred from radio observations (e.g., VLBI measurements).

\subsection{The Evolution of Particle Distribution}
The emission from MWN can be obtained by solving the electron continuity equation
\begin{equation}\label{eq:Ngamma}
\frac{\partial}{\partial t} n_{\mathrm{e}, \gamma}+\frac{\partial}{\partial \gamma_{\mathrm{e}}}\left(\dot{\gamma_{\mathrm{e}}} n_{\mathrm{e}, \gamma}\right)=Q_{\mathrm{e}, \gamma},
\end{equation}
where $n_{\mathrm{e}, \gamma} d \gamma_{\mathrm{e}}$ is the electron number density between Lorentz factors $\gamma_{\mathrm{e}}$ and $\gamma_{\mathrm{e}}+d \gamma_{\mathrm{e}}$. The cooling term $\dot{\gamma}_{\mathrm{e}}$ of electrons includes the radiation cooling (synchrotron radiation and IC scattering) and adiabatic expansion
\begin{equation}\label{eq:dgammadt} 
\dot{\gamma}_{\mathrm{e}}(\gamma_{\mathrm{e}}, t)=\dot{\gamma}_{\mathrm{e,syn}}(\gamma_{\mathrm{e}}, t)+\dot{\gamma}_{\mathrm {e,ic}}(\gamma_{\mathrm{e}}, t)+\dot{\gamma}_{\mathrm {e,ad}}(\gamma_{\mathrm{e}}, t) .
\end{equation}
The energy loss due to synchrotron radiation is \citep{Rybicki1979}
\begin{equation}
\dot{\gamma}_{\mathrm{e,syn}}=-\frac{\sigma_\mathrm{T} B_{\mathrm{n}}^{2} \beta_{\mathrm{e}}^2 \gamma_{\mathrm{e}}^2 }{6 \pi m_{\mathrm{e}} c},
\end{equation}
where $\sigma_\mathrm{T}$ and $m_{\mathrm{e}}$ are the Thomson cross-section and electron mass, respectively. Additionally, the electrons will also lose energy through the IC scattering of synchrotron photons or stellar photons. The detailed SSC and EIC emission process is given in Appendix \ref{sec:rad}. From the constraints provided by VLBI observations, the scale of PRS is estimated to be approximately on the order of parsecs or sub-parsecs \citep{Marcote2017,Bhandari2023}. The EIC energy-loss can be ignored at such a distance. The energy loss due to adiabatic cooling is given by
\begin{equation}
\dot{\gamma}_{\mathrm{e}, \mathrm{ad}}=-\frac{1}{3} \gamma_{\mathrm{e}} \beta_{\mathrm{e}}^2 \frac{d \ln V_{\mathrm{m}}}{d t}=-\gamma_{\mathrm{e}} \beta_{\mathrm{e}}^2 \frac{\dot{R}_{\mathrm{m}}}{R_{\mathrm{m}}}.
\end{equation}

The source term $Q_{\mathrm{e},j}(\gamma_{\mathrm{e}})=\dot{N}_{\mathrm{e},j}\cdot f_j(\gamma_{\mathrm{e}})/V_{\mathrm{m}}$ is the energy distribution of electrons injected into the nebula at the termination shock. For rotation-powered model, the injected electrons are phenomenological descripted as a power-law or a broken power-law distribution. From the study of Galactic PWNe, the electron distribution is given by a broken power-law \citep{Tanaka2010}
\begin{equation}
f_{\mathrm{BPL}}(\gamma_{\mathrm{e}})\propto \begin{cases}\left(\gamma_{\mathrm{e}} / \gamma_{\mathrm{b}}\right)^{-p_1} & \gamma_{\min } \leqslant \gamma_{\mathrm{e}} \leqslant \gamma_{\mathrm{b}} \\ \left(\gamma_{\mathrm{e}} / \gamma_{\mathrm{b}}\right)^{-p_2} & \gamma_{\mathrm{b}} \leqslant \gamma_{\mathrm{e}} \leqslant \gamma_{\mathrm{max}}\end{cases},
\end{equation}
where $\gamma_{\min}, \gamma_\mathrm{b}$, and $\gamma_{\max}$ are the minimum, break, and maximum Lorentz factors of the injected electrons. The power-law indices for low- and high-energy particles are found to be $p_1<2$ and $p_2>2$ \citep{Tanaka2010}. For the magnetic-powered models, the injected electrons obey a Maxwellian-Jüttner distribution \citep{Margalit2018}
\begin{equation}
f_{\mathrm{MJ}}\left(\gamma_{\mathrm{e}},\Theta_{\mathrm{e}}\right)=\frac{\gamma_{\mathrm{e}}^2 \beta_{\mathrm{e}}}{\Theta_{\mathrm{e}} K_2\left(1 / \Theta_{\mathrm{e}}\right)} \exp \left(-\frac{\gamma_{\mathrm{e}}}{\Theta_{\mathrm{e}}}\right),
\end{equation}
where $K_2$ is modified Bessel function of the second kind and  $\Theta_{\mathrm{e}}=k_{\mathrm{B}} T_{\mathrm{e}}/m_{\mathrm{e}} c^2$ is the dimensionless temperature with $k_{\mathrm{B}}$ being Boltzmann’s constant. The total electron number injection rate $\dot{N}_{\mathrm{e},j}$ is given by
\begin{equation}\label{eq:Netot}
\dot{N}_{\mathrm{e},j}=\frac{L_j}{\left(1+\sigma_j\right) \langle\epsilon_j\rangle},
\end{equation}
where $\langle\epsilon_j\rangle$ is the averaged energy per particle carried. For relativistic electron–positron pulsar wind, the averaged energy is $\langle\epsilon_{\mathrm{sd}}\rangle=\langle\gamma_{\mathrm{e}}\rangle m_{\mathrm{e}}c^2$, where the averaged Lorentz factor is $\left\langle\gamma_{\mathrm{e}}\right\rangle=\int \gamma_{\mathrm{e}} f\left(\gamma_{\mathrm{e}}\right) d \gamma_{\mathrm{e}} / \int f\left(\gamma_{\mathrm{e}}\right) d \gamma_{\mathrm{e}}$. The averaged energy of baryon-loaded magnetar giant flares from SGR 1806$-$20 is found to be $\langle\epsilon_{\mathrm{B}}\rangle\sim 0.16-10$ GeV \citep{Granot2006}. Assuming a proton–electron composition of magnetar outflows, the injected electrons are heated to an averaged Lorentz factor \citep{Margalit2018}
\begin{equation}
    \langle\gamma_{\mathrm{e}}\rangle \approx \frac{\langle\epsilon_{\mathrm{B}}\rangle}{2 m_{\mathrm{e}} c^2}\approx 196\left(\frac{\langle\epsilon_{\mathrm{B}}\rangle}{0.2~\mathrm{GeV}}\right).
\end{equation}

The temperature of the injected electrons can be estimated from $\langle\gamma_{\mathrm{e}}\rangle \approx 3 \Theta_{\mathrm{e}}$ for ultrarelativistic temperatures $\Theta_{\mathrm{e}} \gg 1$.


\subsection{MWN emission}  
The synchrotron radiation luminosity at frequency $\nu$ of MWNs is $L_\nu \propto j_\nu/\alpha_\nu\cdot (1-e^{-\tau_{\mathrm{SSA}}})$, where $\tau_{\mathrm{SSA}}=\alpha_\nu R_\mathrm{m}$ is SSA optical depth of the nebula. The synchrotron emissivity $j_\nu$ and absorption coefficients $\alpha_\nu$ are given in Appendix \ref{sec:rad}. The synchrotron luminosity depends on the electron distribution $n_{\gamma}$ and the magnetic field $B_{\mathrm{m}}$, which can be obtained by solving equations (\ref{eq:Ngamma}) and (\ref{eq:Bn}). 

Equation (\ref{eq:Ngamma}) is solved numerically on a discretized electron Lorentz-factor grid. The energy-space flux divergence is evaluated using a fifth-order weighted essentially non-oscillatory (WENO5) reconstruction \citep{Jiang1996}, which provides a high-resolution treatment of the evolving electron distribution. For an explicit time-marching scheme, the allowable time-step size is strictly constrained by the Courant–Friedrichs–Lewy (CFL) condition. To achieve a long-term simulation, an implicit scheme (e.g., the backward Euler method) is applied to the temporal derivative.

\begin{figure}
    \centering
    \includegraphics[width=\linewidth]{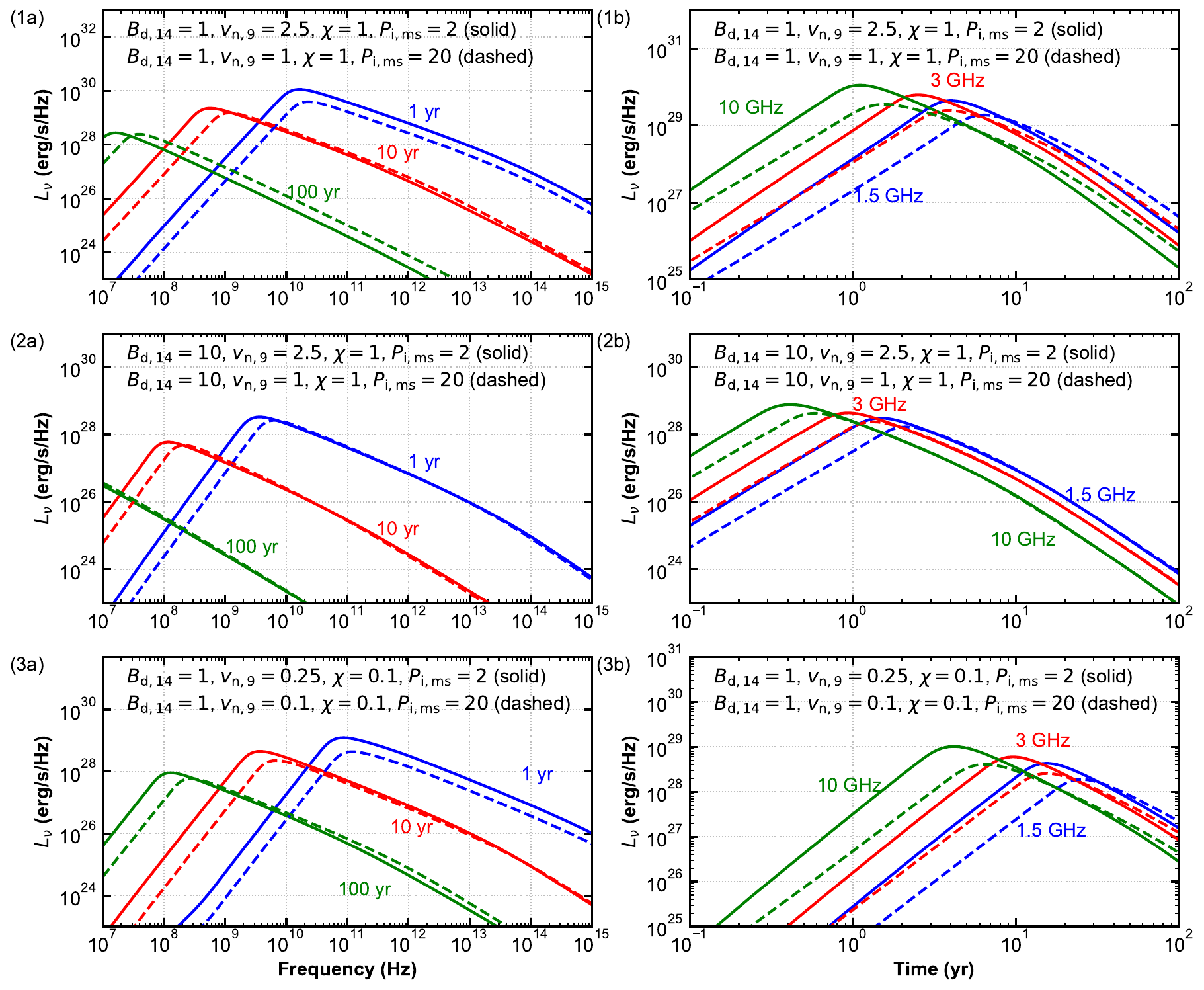}
    \caption{The synchrotron SEDs (panels (1a)-(3a)) and light curves (panels (1b)-(3b)) of the rotation-powered nebula. The left  panels show the spectrum for ages $t_{\mathrm{age}} = 1$ yr (blue lines), 10 yr (red lines), and 100 yr (green lines). For the initial spin period $P_{\mathrm{i}}=2$ ms, the final accelerated expanding velocity of the nebula is $v_{\mathrm{n}}=2.5\times 10^9$ cm/s (solid line). For the cases of $P_{\mathrm{i}}=20$ ms, the nebula is not accelerated, and the expanding velocity is $v_{\mathrm{n}}=v_{\mathrm{ej}}=1\times 10^9$ cm/s (dashed lines). The right panels show the light curves at frequencies $\nu=1.5$ GHz (blue lines), $\nu=3$ GHz (red lines), and $\nu=10$ GHz (green lines). \textbf{Panel (1a-1b):} for $B_{\mathrm{dip}}=10^{14}$ G and $\chi=1$. \textbf{Panel (2a-2b):} for $B_{\mathrm{dip}}=10^{15}$ G and $\chi=1$. For the case of a higher dipolar magnetic field, the synchrotron luminosity is 1–2 orders of magnitude lower. The shorter spin-down timescale means the faster decay of the luminosity. \textbf{Panel (3a-3b):} for $B_{\mathrm{dip}}=10^{14}$ G and $\chi=0.1$. A smaller radiation region results in higher peak frequency and lower luminosity synchrotron radiation at the same age. }
    \label{fig:rot}
\end{figure}

In this section, we will present some numerical results to illustrate the impact of the model parameters. In our calculations, the typical SNe parameters $M_{\mathrm{ej}} = 1M_\odot$ and $E_{\mathrm{sn}} = 10^{51}$ erg are employed. In the left panels of Figure \ref{fig:rot}, the spectral energy distribution (SED) of the rotation-powered nebula is depicted for ages $t_{\mathrm{age}} = 1$ yr (blue lines), 10 yr (red lines), and 100 yr (green lines). The corresponding radio light curves at frequencies $\nu=1.5$ GHz (blue lines), $\nu=3$ GHz (red lines), and $\nu=10$ GHz (green lines) are presented in the right panels of Figure \ref{fig:rot}. 

As shown in Figure \ref{fig:rot}(1a), we present the SED for the rotation-powered nebula with $B_{\mathrm{d}} = 10^{14}$ G and $\chi=1$. For the initial spin period $P_{\mathrm{i}}=2$ ms, the final accelerated expanding velocity of the nebula is $v_{\mathrm{n}}=2.5\times 10^9$ cm/s (solid line). Even for millisecond magnetars, rapid spin-down limits the luminosity of PRS to only reach above $10^{29}$ erg/s/Hz when they are extremely young (e.g., $t_{\mathrm{age}} < 10 $ yr, see Figure \ref{fig:rot}(1b)). However, the SNR is expected to be optically thick for radio signals at an age $\lesssim 10$ yr due to the free-free absorption of SN ejecta \citep{Metzger2017,Zhao2021}. For the cases of $P_{\mathrm{i}}=20$ ms, the nebula is not accelerated, and the expanding velocity is $v_{\mathrm{n}}=v_{\mathrm{ej}}=1\times 10^9$ cm/s (dashed lines). For the case of a higher dipolar magnetic field (e.g, $B_{\mathrm{d}}=10^{15}$ G), the synchrotron luminosity is 1–2 orders of magnitude lower (Figure \ref{fig:rot}(2a)). The shorter spin-down timescale means the faster decay of the luminosity (see Figure \ref{fig:rot}(1b) and \ref{fig:rot}(2b)). Considering that the radiation region can be significantly smaller than the nebula radius, we also show the SED and light curves for $\chi=0.1$ in Figure \ref{fig:rot}(3a) and \ref{fig:rot}(3b). At the same age, a smaller radiation region results in higher peak frequency and lower luminosity synchrotron radiation. 

\begin{figure}
    \centering
\includegraphics[width=\linewidth]{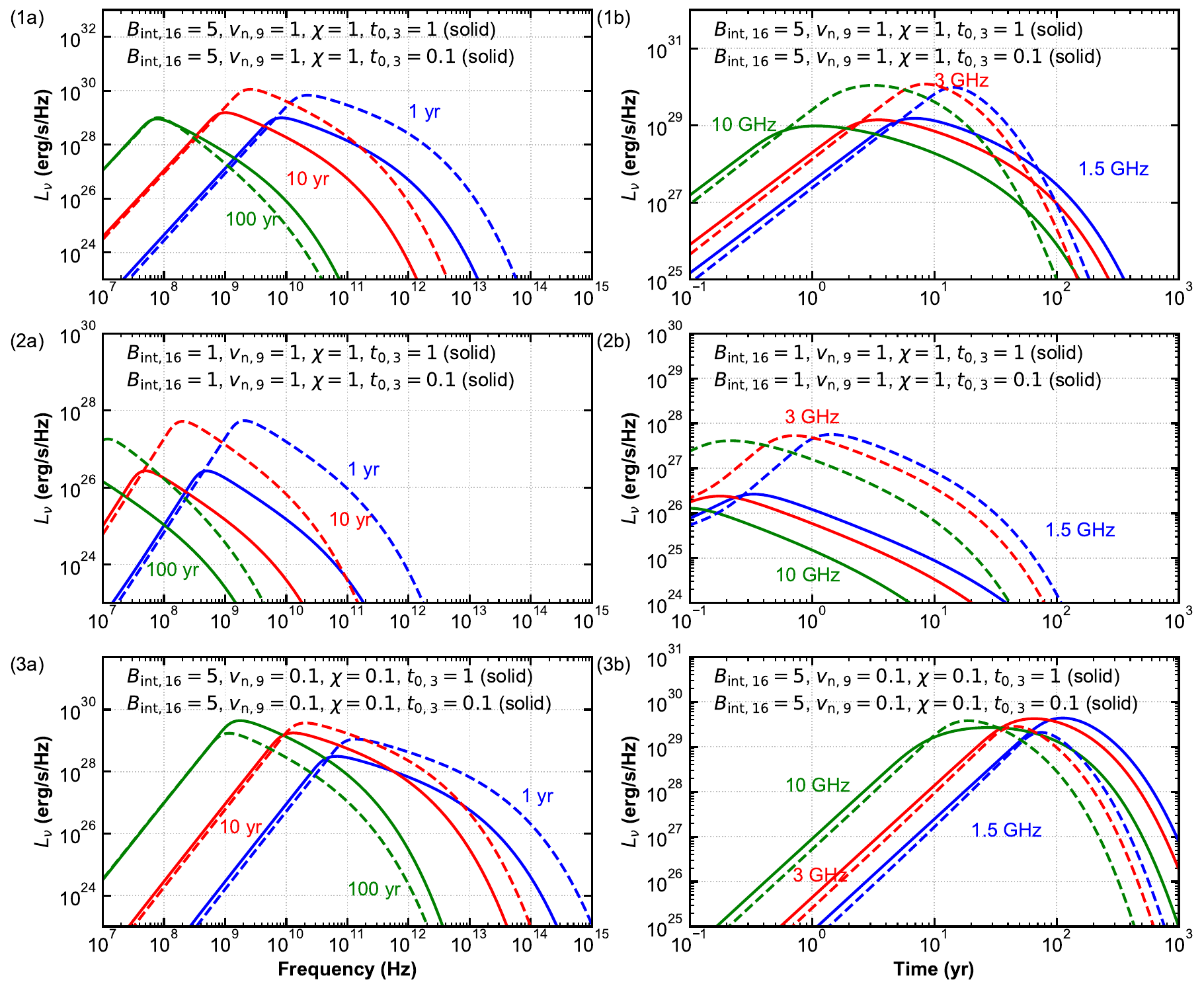}
    \caption{Same as Figure \ref{fig:rot}, but for the magnetic-powered model. Compared to rotational energy, the internal magnetic energy of magnetars has a longer decay timescale, which allows the luminosity of MWNs to remain stable over a period of $1-100$ yr.}
    \label{fig:mag}
\end{figure}

The numerical results of the SED and light curves of the magnetic-powered MWN are shown in Figure \ref{fig:mag}. To explain the bright PRS (with luminosity of $\nu L_{\nu}\sim 10^{38}-10^{39}$ erg/s), the magnetar's internal magnetic field must be greater than $10^{16}$ G. Here, we show the energy spectrum evolution for the cases with $B_{\mathrm{int}}=5\times 10^{16}$ G (see Figure \ref{fig:mag}(1a) and \ref{fig:mag}(1b)) and $B_{\mathrm{int}}=10^{16}$ G (see Figure \ref{fig:mag}(2a) and \ref{fig:mag}(2b)). The magnetar's total internal magnetic energy usually does not exceed the supernova explosion energy (Equation (\ref{eq:EBint})). Therefore, the injection of magnetic energy will not cause effective acceleration of the nebula. Compared to rotational energy, the internal magnetic energy of magnetars has a longer decay timescale, which allows the luminosity of MWNs to remain stable over a period of $1-100$ yr (right panels of Figure \ref{fig:mag}). The Maxwell–Jüttner distribution has an exponentially suppressed high-energy tail $f_{\mathrm{MJ}}(\gamma_{\mathrm{e}}) \propto \gamma_{\mathrm{e}}^2 \exp \left(-\gamma_{\mathrm{e}} / \Theta_{\mathrm{e}}\right)$. In contrast, the extended non-thermal tail of the broken power-law distribution ($f_{\mathrm{BPL}}(\gamma_{\mathrm{e}})  \propto \gamma_{\mathrm{e}}^{-p_2}$) produces broader spectral components and more extended high-frequency emission.

\section{The emission from the wind interactions}\label{sec:binary}
\subsection{Electron distribution in bow shocks}
The evolution of the electron distribution is determined by the continuity equation (\ref{eq:Ngamma}). However, the cooling timescale is usually much smaller than the orbital period of the massive binary systems. The time-dependent term in equation (\ref{eq:Ngamma}) can be neglected ($\partial n_{\gamma}/\partial t=0$). Under the above steady-state approximation, the solution of the continuity equation is
\begin{equation}\label{eq:n_gamma_steady}
n_{\mathrm{e}}(\gamma_{\mathrm{e}})=\frac{1}{|\dot{\gamma_{\mathrm{e}}}|} \int Q\left(\gamma_{\mathrm{e}}\right) \mathrm{d} \gamma_{\mathrm{e}} .
\end{equation}

The unshocked electrons in the pulsar wind will be accelerated to ultra-relativistic speeds in the termination shock. The injected electron is usually assumed to follow a power-law distribution $Q\left(\gamma_{\mathrm{e}}\right) \sim \gamma_{\mathrm{e}}^{-p}$ for $\gamma_{\mathrm{e}, \min }<\gamma_{\mathrm{e}}<\gamma_{\mathrm{e}, \max }$, where $p$ is the electron spectrum index. The minimum Lorentz factor of the shocked electrons is determined by \citep{Kirk1999}
\begin{equation}
\gamma_{\mathrm{e}, \min }=\Gamma_{\mathrm{w}} \frac{p-2}{p-1},
\end{equation}
where $\Gamma_{\mathrm{w}}$ is the Lorentz factor of the pre-shock pulsar wind. The pulsar wind is primarily composed of ultra-relativistic ($\Gamma_{\mathrm{w}}\sim 10^5-10^6$) electron-positron pairs. The balance between the acceleration and the cooling process determines the maximum Lorentz factor
\begin{equation}
\gamma_{\mathrm{e}, \max }=\sqrt{\frac{6 \pi e}{\sigma_{\mathrm{T}} B_\mathrm{b}(1+Y)}},
\end{equation}
where the Compton parameter $Y$ is given in Section \ref{sec:ic}, considering both SSC and EIC processes. The cooling timescale of the electrons is $t_{\text {cool }}=\gamma_{\mathrm{e}} m_{\mathrm{e}} c^2 / P_{\mathrm{rad}}$, where $P_{\mathrm{rad}}=(4 / 3) \sigma_{\mathrm{T}} c \gamma_{\mathrm{e}}^2\left(B_\mathrm{b}^2 / 8 \pi\right)(1+Y)$ is the radiation power. The cooling is effective for the electrons with the Lorentz factor above the critical Lorentz factor $\gamma_{\mathrm{e}, \mathrm{c}}$, given by equating the cooling timescale to the dynamic timescale \citep{Sari1998}
\begin{equation}
\gamma_{\mathrm{e}, \mathrm{c}}=\frac{6 \pi m_{\mathrm{e}} c}{\sigma_{\mathrm{T}} B_\mathrm{b}^2 \tau_{\mathrm{dyn}}(1+Y)} .
\end{equation}

For the power-law electron injection spectrum of the rotational-energy-powered models, the electron distribution is
\begin{equation}
n_\gamma \propto \begin{cases}\gamma_{\mathrm{e}}^{-2}, & \gamma_{\mathrm{e}, \mathrm{c}} \leq \gamma_{\mathrm{e}}<\gamma_{\mathrm{e}, \min } \\ \gamma_{\mathrm{e}}^{-(p+1)}, & \gamma_{\mathrm{e}, \min } \leq \gamma_{\mathrm{e}} \leq \gamma_{\mathrm{e}, \max }\end{cases}
\end{equation}
in the fast-cooling case and 
\begin{equation}
n_\gamma \propto \begin{cases}\gamma_{\mathrm{e}}^{-p}, & \gamma_{\mathrm{e}, \min } \leq \gamma_{\mathrm{e}} \leq \gamma_{\mathrm{e}, \mathrm{c}} \\ \gamma_{\mathrm{e}, \mathrm{c}}\gamma_{\mathrm{e}}^{-(p+1)}, & \gamma_{\mathrm{e}, \mathrm{c}}<\gamma_{\mathrm{e}} \leq \gamma_{\mathrm{e}, \max }\end{cases}
\end{equation}
in the slow-cooling case. The cooled electron distribution for power-law electron injection can be represented by simple broken power-laws. However, for Maxwell–Jüttner electron injection, the standard slow- and fast-cooling prescriptions do not offer much convenience in the calculation. We therefore obtain the cooled electron distribution numerically using equation (\ref{eq:n_gamma_steady}) for magnetic-powered models.

The magnetic field in the bow shock is given by equation (\ref{eq:Bn}). If we neglect the effect of adiabatic expansion, the magnetic field is
\begin{equation}
B_\mathrm{b}=\left[\frac{8 \pi \sigma_j\dot{E}_j \tau_{\mathrm{d}}}{V_{\mathrm{b}} (1+\sigma_j)}\right]^{1 / 2},
\end{equation}
where $V_{\mathrm{b}}=4\pi r_{\mathrm{s}}^2 D$ is the volume of the emission region with $D=\frac{1}{8} \min \left(r_{\mathrm{s}}, R_{\mathrm{s}}\right)$ being thickness of the shocked region \citep{Luo1990}. The total injected electron number is $N_{\mathrm{e}}=\dot{N}_{\mathrm{e},j}\cdot \tau_{\mathrm{d}}$, where the injection rate is given in equation (\ref{eq:Netot}).The dynamic timescale of electrons is $\tau_{\mathrm{d}}\simeq r_{\mathrm{s}}/\beta_j c$. The similar  result $B_\mathrm{b}\propto (\dot{E}_j/r_{\mathrm{s}}^2)^{1/2}$ is also presented in \cite{Kennel1984a,Kennel1984b}. However, we define the magnetization parameter of the injected outflows, rather than that of the pre-shocked pulsar wind used in the KC model.

\subsection{The Multiwavelength Radiation from Wind Interaction}
The synchrotron and IC radiation power at frequency $\nu$ from the bow shock is 
\begin{equation}
L_\nu^{\mathrm{tot}}=L_\nu^{\mathrm{syn}} \frac{1-e^{-\tau_{\mathrm{SSA}}}}{\tau_{\mathrm{SSA}}}+(L_\nu^{\mathrm{SSC}}+L_\nu^{\mathrm{EIC}}) e^{-\tau_{\gamma \gamma}},
\end{equation}
where $\tau_{\mathrm{SSA}}\sim \alpha_{\nu}D$ and $\tau_{\gamma \gamma}\sim n_\star \sigma_{\gamma \gamma} D$ is the optical depth for SSA and $\gamma \gamma$ annihilation (see Appendix \ref{sec:rad}). The SSA and $\gamma \gamma$ annihilation affect low-energy photons of synchrotron radiation and high-energy photons of IC adiation, respectively. In the binary system, the absorption of gamma-ray photons via $\gamma \gamma$ annihilation mainly comes from the stellar photon field. The photon density at the radial distance $r$ away from the star is 
\begin{equation}
n_{\star}(r)=\frac{L_{\star}}{4 \pi r^2 c\left\langle\epsilon_{\star}\right\rangle},
\end{equation}
where $L_\star$ is the stellar luminosity. The characteristic stellar photon energy is $\left\langle\epsilon_{\star}\right\rangle\sim 2.82 k_{\mathrm{B}}T_\star=2.4 T_{\star,4}\mathrm{~eV}$. Adopting the $\gamma \gamma$ annihilation cross section $\sigma_{\gamma \gamma}\lesssim \sigma_{\mathrm{T}}/5$,
the optical depth at the orbital separation ($r\sim d, D\sim d/8$) can be estimated as 
\begin{equation}
    \begin{aligned}
 \tau_{\gamma \gamma}&\sim n_\star \sigma_{\gamma \gamma} D=\frac{L_{\star}\sigma_{\gamma \gamma}}{32 \pi d c\left\langle\epsilon_{\star}\right\rangle}\\
 & \lesssim 0.061 L_{\star,38}\left(\frac{d}{\mathrm{AU} }\right)^{-1}\left(\frac{\left\langle\epsilon_{\star}\right\rangle}{3~\mathrm{eV} }\right)^{-1}.
    \end{aligned}
\end{equation}
Based on the study of long-term DM and RM variations \citep{Wang2022,Zhao2023,Wang2025}, it is suggested that HMXBs containing the active FRBs that we are interested in tend to be long-period systems ($P_\mathrm{orb}>100$ d and $d>1$ AU). The absorption of gamma-ray photons via $\gamma \gamma$ annihilation can be ignored in our calculations. 

\begin{figure}
    \centering
\includegraphics[width=\linewidth]{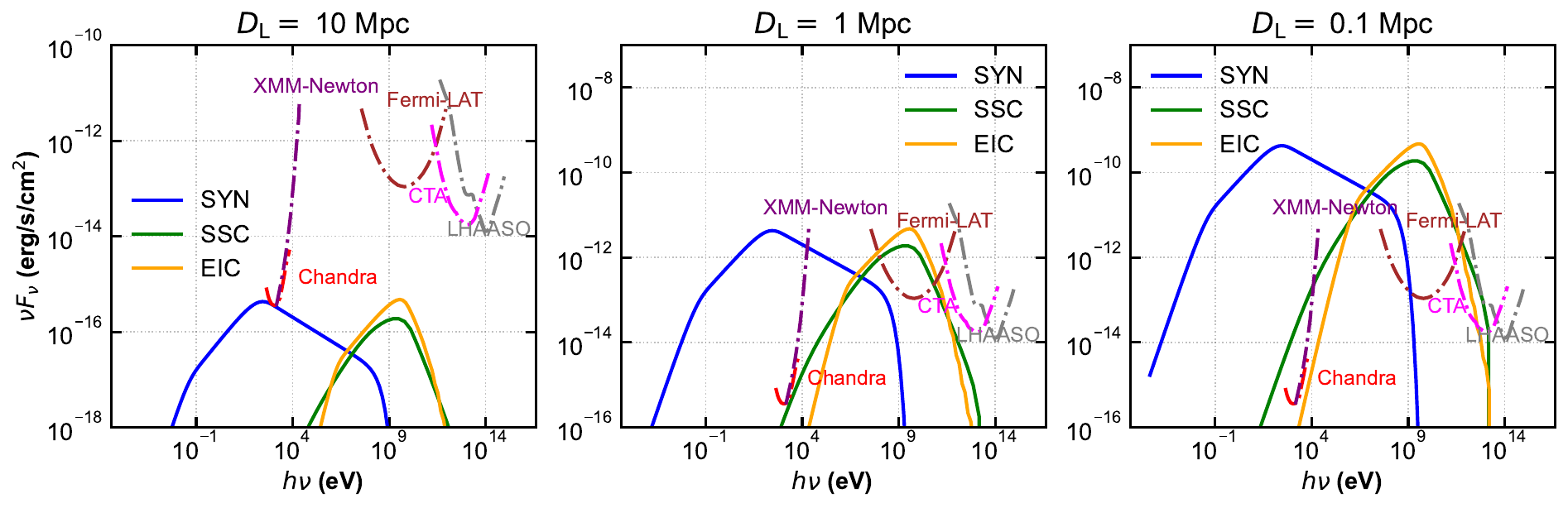}
    \caption{The spectra of synchrotron radiation (blue lines), SSC (green lines), and EIC  scattering radiation (orange lines) from the bow shock of rotation-powered models ($\Gamma_{\mathrm{w}}=10^5$ and $\sigma=0.01$). For simplicity, we assume that the luminosity of the central engine is constant ($\dot{E}=10^{40}$ erg/s) and that the binary systems are in circular orbits. The period of the binary system is $P=100$ d. The typical massive star and stellar wind parameters are used in our calculation: $M_{\odot}=30M_{\odot}, R_{\odot}=10R_{\odot},T_{\mathrm{eff}}=2\times 10^4$ K, $\dot{M}=5\times 10^{-7} M_{\odot}$ yr$^{-1}$ and $v_{\mathrm{w}}=3\times 10^{8}$ cm/s. The synchrotron radiation of nonthermal electrons in the bow shock is mainly in the keV X-ray band and can be detected by \textit{Chandra} and \textit{XMM-Newton} at the luminosity distance of $D_{\mathrm{L}}\sim 10$ Mpc (left panel). The SSC and EIC scattering radiation is mainly at the GeV band, which can be detected by \textit{Fermi-LAT} at the luminosity distance of $D_{\mathrm{L}}\sim 1$ Mpc (middle panel). At a closer distance (e.g., $D_{\mathrm{L}}\lesssim 0.1$ Mpc), the TeV $\gamma$-ray emission from the IC scattering process can be detected by CTA and LHAASO (right panel). }
    \label{fig:shock1}
\end{figure}

\begin{figure}
    \centering
\includegraphics[width=\linewidth]{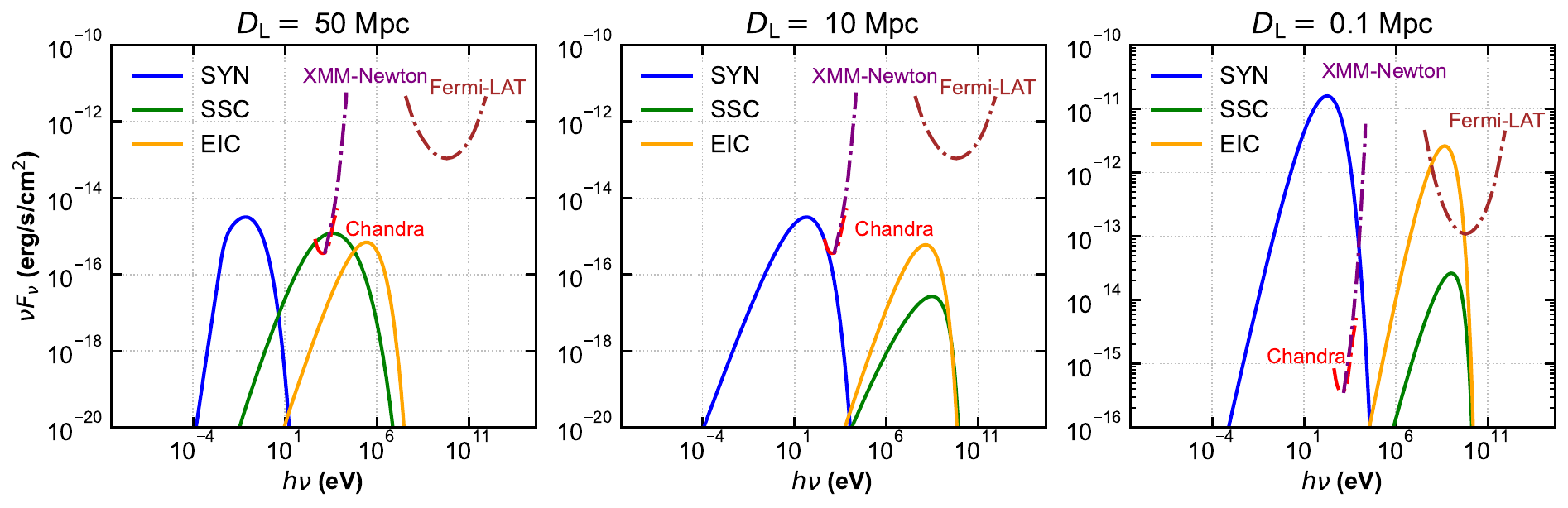}
    \caption{Same as Figure (\ref{fig:shock1}) but for magnetic-powered models. The magnetization parameter $\sigma =0.1$ and the magnetar outflow velocity $\beta=0.7$ is used \citep{Granot2006}. The SEDs of synchrotron radiation, SSC, and EIC scattering radiation of injected electrons with temperatures $k_{\mathrm{B}}T_{\mathrm{e}}\simeq 0.03,0.83,1.67$ GeV are shown in the left, middle and right panels. The synchrotron radiation of low-temperature electrons is mainly in the radio bands, which can account for some faint PRSs associated with FRBs. The SSC and EIC process mainly produces the emission at the $\sim$keV$-\sim$GeV band. The X-ray emission of the bow shock can be detected by \textit{Chandra} and \textit{XMM-Newton} at the luminosity distance of $D_{\mathrm{L}}\sim 50$ Mpc. The X-ray emission from the synchrotron process of high-temperature electrons can be detected at a luminosity distance of $D_{\mathrm{L}}\sim 10$ Mpc, and the GeV radiation of EIC process can be detected by \textit{Fermi-LAT} at a luminosity distance of $D_{\mathrm{L}}\sim 0.1$ Mpc. }
    \label{fig:shock2}
\end{figure}

The SEDs of the rotation-powered and magnetic-powered bow shock are shown in Figure \ref{fig:shock1} and Figure \ref{fig:shock2}, respectively. For simplicity, we assume that the luminosity of the central engine is constant ($\dot{E}=10^{40}$ erg/s) and that the binary systems are in circular orbits. The period of the binary system is $P=100$ d. The typical massive star and stellar wind parameters are used in our calculation: $M_{\odot}=30M_{\odot}, R_{\odot}=10R_{\odot},T_\star=2\times 10^4$ K, $\dot{M}=5\times 10^{-7} M_{\odot}$ yr$^{-1}$ and $v_{\mathrm{w}}=3\times 10^{8}$ cm/s. The spectra of synchrotron radiation, SSC, and EIC scattering radiation are shown in blue, green, and orange solid lines, respectively. 

For rotation-powered models, the pulsar wind Lorentz factors and the magnetization parameter were adopted as $\Gamma=10^5$ and $\sigma=0.01$, which is consistent with the Galactic gamma-ray binaries \citep{Tavani1997,Dubus2006}. The synchrotron radiation of nonthermal electrons in the bow shock is mainly in the keV X-ray band and can be detected by \textit{Chandra} and \textit{XMM-Newton} at the luminosity distance of $D_{\mathrm{L}}\sim 10$ Mpc (left panel of Figure \ref{fig:shock1}). The sensitivity curves of \textit{Chandra} and \textit{XMM-Newton} with an exposure time of $10^3$ s are shown in red and purple dash-dotted lines, respectively. The SSC and EIC scattering radiation is mainly at the GeV band, which can be detected by \textit{Fermi-LAT} at the luminosity distance of $D_{\mathrm{L}}\sim 1$ Mpc (middle panel of Figure \ref{fig:shock1}). The sensitivity curves of \textit{Fermi-LAT} are shown in brown dashed–dotted lines \footnote{The sensitivity curves of Fermi are taken from \url{https://www.slac.stanford.edu/exp/glast/groups/canda/lat_Performance.htm}.}. At a closer distance (e.g., $D_{\mathrm{L}}\lesssim 0.1$ Mpc), the TeV $\gamma$-ray emission from the IC scattering process can be detected by CTA and LHAASO. The magenta and gray dashed–dotted lines indicate the sensitivity curves of CTA \citep{CherenkovTelescopeArrayConsortium2019} and LHAASO \citep{Cao2019}.

For magnetic-powered models, the magnetization parameter $\sigma =0.1$ and the magnetar outflow velocity $\beta=0.7$ is used \citep{Granot2006}. The SEDs of synchrotron radiation, SSC, and EIC scattering radiation of injected electrons with temperatures $k_{\mathrm{B}}T_{\mathrm{e}}\simeq 0.03,0.83,1.67$ GeV are shown in the left, middle and right panels of Figure \ref{fig:shock2}. The synchrotron radiation of low-temperature electrons is mainly in the radio bands, which can account for some faint PRSs associated with FRBs. The SSC and EIC process mainly produces the emission at the $\sim$keV$-\sim$GeV band. The X-ray emission of the bow shock can be detected by \textit{Chandra} and \textit{XMM-Newton} at the luminosity distance of $D_{\mathrm{L}}\sim 50$ Mpc. The X-ray emission from the synchrotron process of high-temperature electrons can be detected at a luminosity distance of $D_{\mathrm{L}}\sim 10$ Mpc, and the GeV radiation of EIC process can be detected by \textit{Fermi-LAT} at a luminosity distance of $D_{\mathrm{L}}\sim 0.1$ Mpc.


\subsection{Orbital-dependent Multiwavelength Radiation}
The synchrotron and SSC radiation from the bow shock depends on the standoff distance of the wind interaction, which is highly dependent on the orbital phase. The EIC process is related not only to the wind interaction but also to the orbital-dependent scattering angle. Adopting the same orbital geometry as in our previous work \citep{Zhao2023,Wang2025}, the scattering angle is given by $\mu=\boldsymbol{e}_{\mathrm{obs}} \cdot \boldsymbol{e}_{\mathrm{m}}$, where $
\boldsymbol{e}_{\mathrm{obs}}=\left(\sin \theta_{\mathrm{o}} \cos \phi_{\mathrm{o}}, \sin \theta_{\mathrm{o}} \sin \phi_{\mathrm{o}}, \cos \theta_{\mathrm{o}}\right)
$ and $
\boldsymbol{e}_{\mathrm{m}}=(\cos \phi, \sin \phi, 0)
$ is the unit vector describing the direction of the observer and the magnetar, respectively. In this section, we present the multiwavelength light curves that depend on the orbital parameters.  Other parameters are the same as Figure \ref{fig:shock1}.

\begin{figure}
    \centering
    \includegraphics[width=\linewidth]{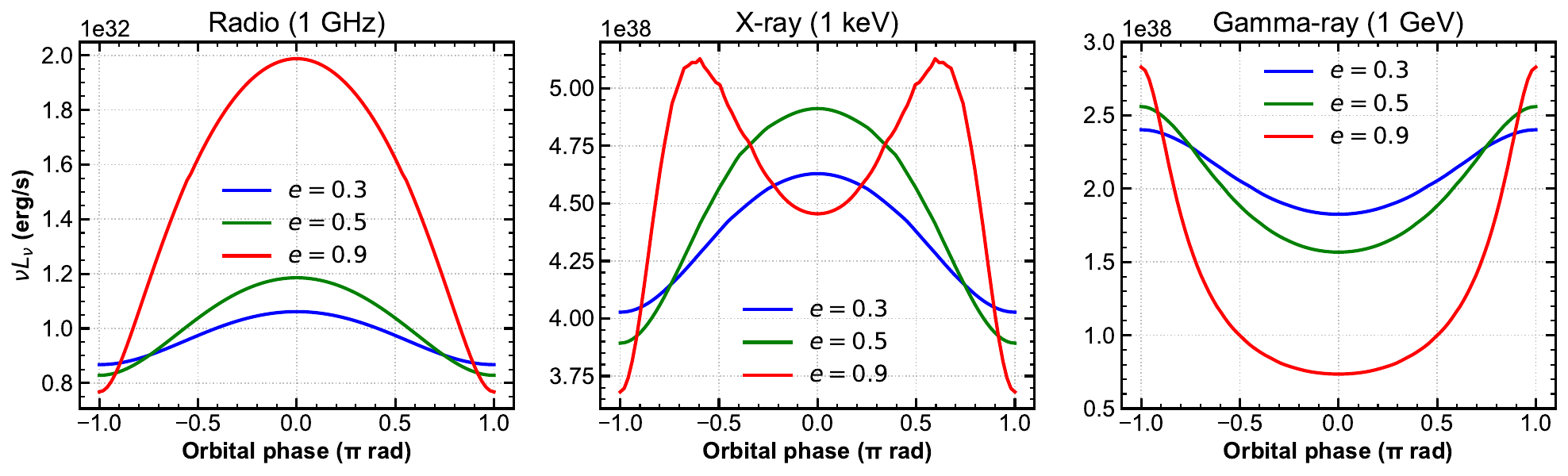}
    \caption{The synchrotron GHz radio emission (left panel), keV X-ray emission (middle panel) and SSC GeV $\gamma-$ray emission (right panel) at different orbital phase. The red, green and red solid lines illustrate the case of eccentricity $e=0.3,0.5,0.9$. Other parameters are the same as Figure (\ref{fig:shock1}). The synchrotron luminosity originating from the bow shock can vary significantly at different orbital phases in an eccentric orbit, with higher eccentricity resulting in larger luminosity variations.}
    \label{fig:syn_ssc_bin}
\end{figure}

\begin{figure}
    \centering
    \includegraphics[width=\linewidth]{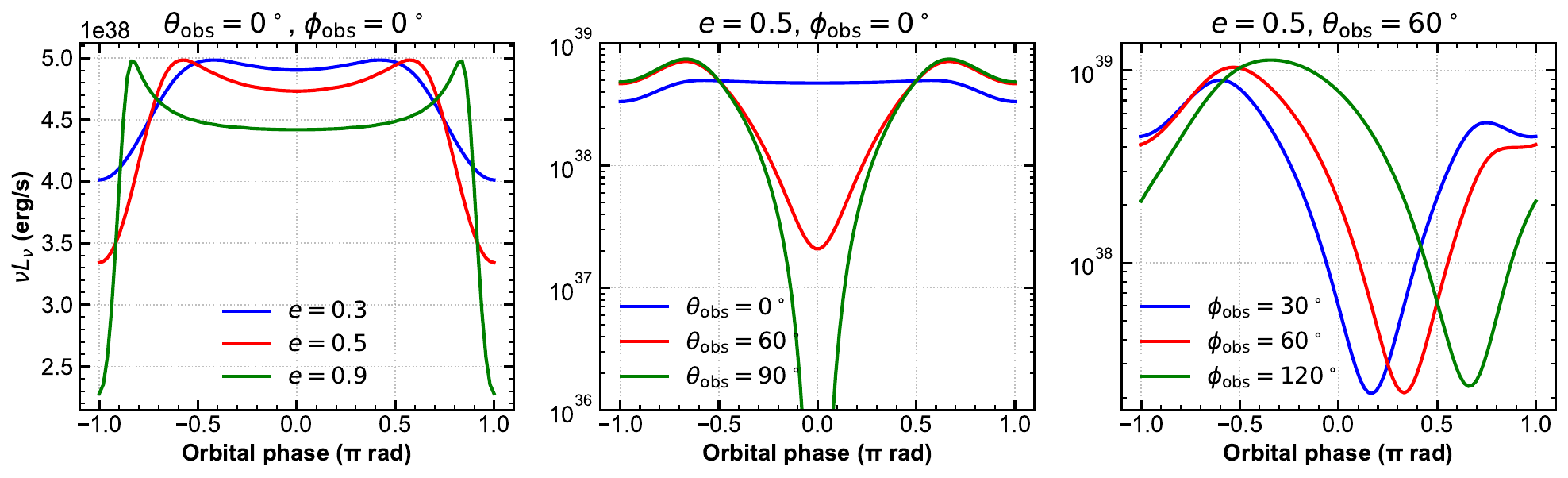}
    \caption{The EIC radiation GeV light curves for different orbital geometry parameters. Other parameters are the same as Figure (\ref{fig:shock1}). \textbf{Left panel:}the blue, green, and red solid lines illustrate the cases of eccentricity $e=0.3, 0.5, 0.9$. The observation angle $(\theta_{\mathrm{o}},\phi_{\mathrm{o}})=(0^\circ,0^\circ)$ is used. \textbf{Middle panel:} The GeV EIC radiation of  observation inclination angles with $\theta_{\mathrm{o}}=0^\circ$ (blue line), $60^\circ$ (green line) and $90^\circ$ (red line). In the situation where the system is observed face-on ($\theta_{\mathrm{o}}=0^\circ$), there is minimal variation in luminosity with respect to the orbital phase. However, when observed edge-on ($\theta_{\mathrm{o}}=90^\circ$), the luminosity may fluctuate significantly, spanning several orders of magnitude as the orbital phase changes. \textbf{Right panel:} the luminosity for true anomaly angles of the observer with $\phi_{\mathrm{o}}=30^\circ$ (blue line), $\phi_{\mathrm{o}}=60^\circ$ (green line) and $\phi_{\mathrm{o}}=120^\circ$ (red line). The observer's true anomaly angle determines the orbital phase of the luminosity extremum: it is minimum at $\phi-\phi_{\mathrm{o}}=0$ and maximum at $\phi-\phi_{\mathrm{o}}=-\pi$.}
    \label{fig:eic_bin}
\end{figure}

The synchrotron GHz radio, keV X-ray emission and SSC GeV $\gamma-$ray emission at different orbital phases are shown in the left, middle and right panels in Figure \ref{fig:syn_ssc_bin}. The blue, green, and red solid lines illustrate the cases of eccentricity $e=0.3, 0.5, 0.9$.  The EIC radiation GeV light curves for different orbital geometry parameters are shown in Figure \ref{fig:eic_bin}. Other parameters are the same as Figure \ref{fig:shock1}.
In the left panel of Figure \ref{fig:eic_bin}, the blue, green, and red solid lines illustrate the cases of eccentricity $e=0.3, 0.5, 0.9$. The observation angle $(\theta_{\mathrm{o}},\phi_{\mathrm{o}})=(0^\circ,0^\circ)$ is used. The luminosity of the synchrotron and IC radiation originating from the bow shock can vary significantly at different orbital phases in an eccentric orbit, with higher eccentricity resulting in greater luminosity variations.

The middle panel of Figure \ref{fig:eic_bin} shows the GeV EIC radiation of different observation inclination angles. The blue, green, and red solid lines represent the angles of $\theta_{\mathrm{o}}=0^\circ, 60^\circ,90^\circ$. In the situation where the system is observed face-on ($\theta_{\mathrm{o}}=0^\circ$), there is minimal variation in luminosity with respect to the orbital phase. However, when observed edge-on ($\theta_{\mathrm{o}}=90^\circ$), the luminosity may fluctuate significantly, spanning several orders of magnitude as the orbital phase changes. The right panel of Figure \ref{fig:eic_bin} shows the luminosity for different true anomaly angles of the observer. The blue, green, and red solid lines represent the angles of $\phi_{\mathrm{o}}=30^\circ, 60^\circ,120^\circ$. The observer's true anomaly angle determines the orbital phase of the luminosity extremum: it is minimum at $\phi-\phi_{\mathrm{o}}=0$ and maximum at $\phi-\phi_{\mathrm{o}}=-\pi$.

\section{Spectral and Temporal Properties of PRS}\label{sec:results}
The synchrotron radiation from the nebula can account for the observed flux density of PRSs
\begin{equation}
F_{\nu_{\mathrm{obs}}}=\frac{(1+z)L_{\nu_{\mathrm{s}}}}{4 \pi D_{\mathrm{L}}^2} ,      
\end{equation}
where $z$ is the redshift,  $D_{\mathrm{L}}$ is the luminosity distance, $\nu_{\mathrm{s}}=(1+z) \nu_{\mathrm{obs}}$ is the emitting frequency in the source frame, and $\nu_{\mathrm{obs}}$ is the observed frequency. The summary of the radio observations for the different observation epochs and telescopes of PRSs is listed in Table \ref{tab:prs_obs}. This section presents the outcomes of spectral and temporal fitting analyses applied to observed PRSs. The parameters utilized in our calculations can be classified accordingly, with specified numerical values indicating the use of fixed values.
\begin{itemize}
    \item The type of central engine: For the rotation-powered model, the parameters are the dipolar magnetic field $B_{\mathrm{d}}$ and the initial spin period $P_{\mathrm{i}}$. For the magnetic-powered model, the parameters are the initial internal magnetic field $B_{\mathrm{int}}$, the fiducial magnetic field decay timescale $t_0$ and the magnetic field decay index $\alpha_{\mathrm{B}}=1$ \citep{Rahaman2025}.
    \item Microphysical parameters: For the rotation-powered model, parameters are the magnetization parameters $\sigma$, the minimum $\gamma_{\min}$, break $\gamma_\mathrm{b}$, and maximum Lorentz factors $\gamma_{\max }$ of the injected electrons, and the electron distribution index $p_1=1.5$ and $p_2=2.5$. For the magnetic-powered model, the injected electrons are modeled with a relativistic Maxwellian distribution with the temperature $\Theta$.
    \item Dynamical parameters: Assuming a constant expanding velocity, the dynamical radius of the nebula is $R_{\mathrm{n}}\sim v_{\mathrm{n}} t$. The size of the magnetized emission region is $\sim \chi R_{\mathrm{n}}$.
    \item The age of the source $t_{\mathrm{age}}$. It is worth noting that observations of the same source may continue for several years. For ease of comparison with source ages estimated by other methods (e.g, the study of the DM and RM, see \citealt{Wang2025}), the age in this work refers to the age at the time of source discovery.
\end{itemize}

The parameters of fitted spectrum and light curves for the four PRSs, such as FRB 20121102A, FRB 20190520B, FRB 20201124A, and FRB 20240114A are listed in Table \ref{tab:results}. Because repeatedly solving Equation (\ref{eq:Ngamma}) is computationally expensive, we do not perform an exhaustive search over the entire multidimensional parameter space. The parameter sets reported in Table \ref{tab:results} are fitted by hand, which are those that reproduce the main observational features while satisfying the adopted physical and observational constraints. Because several parameters have partially degenerate effects on the electron distribution and radiative output, the values in Table \ref{tab:results} should be regarded as representative acceptable parameter sets rather than unique or precisely determined measurements.

\subsection{Radio Radiation from the nebula}\label{sec:prs_fitting}
\subsubsection{FRB 20121102A and FRB 20190520B}
\begin{figure}
    \centering
    \includegraphics[width=\linewidth]{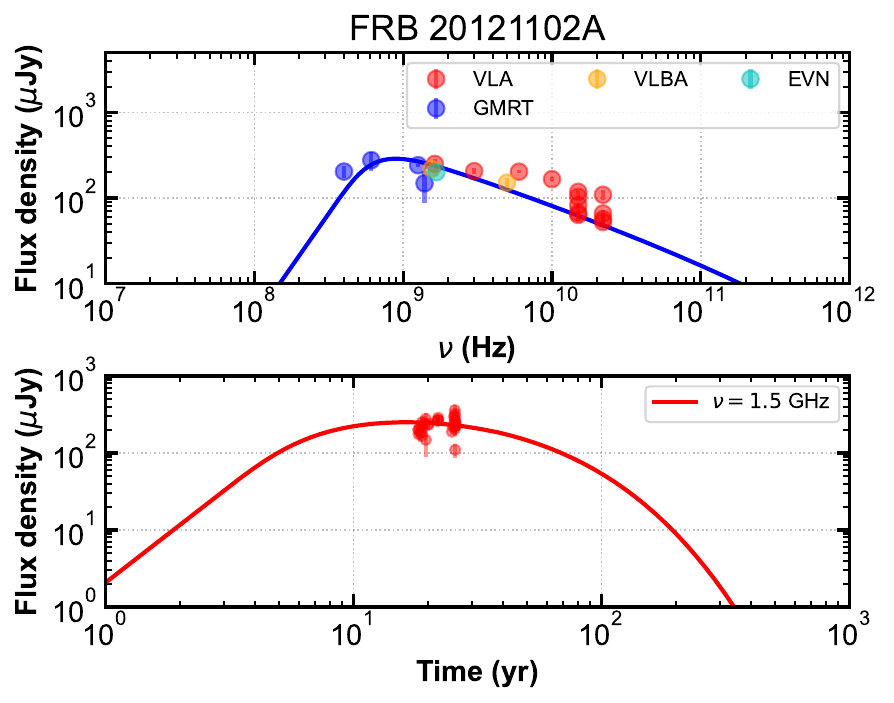}
    \caption{The fitting results of SEDs (top panel) and light curves (bottom panel) of PRS associated with FRB 20121102A. The observation data for the different observation epochs are listed in Table \ref{tab:prs_obs}. The temporal $L$-band (1.3-1.7 GHz) observations of PRS associated with FRB 20121102A are taken from Table 2 in \cite{Bhardwaj2025}, and the light curve with $\nu=1.5$ GHz is shown in red lines in the bottom panel. A young magnetar with $t_{\mathrm{age}}=15$ yr, $B_{\mathrm{int}}=2.7\times 10^{16}$ G and $t_0=100$ yr can explain the observed PRS for FRB 20121102A. Other parameters are listed in Table \ref{tab:results}. The age obtained by fitting the PRS flux is consistent with the conclusions of previous work, which are roughly a dozen years old, whether through the analysis of the PRS flux \citep{Margalit2018,Zhao2021,Bhattacharya2025} and DM/RM variations \citep{Zhao2021a,Wang2025}. The physical size of the emission region is $R_{\mathrm{m}}\sim \chi v_{\mathrm{n}}t=0.15$ pc for FRB 20121102A, which satisfies the given constraint of $<0.7$ pc of VLBI observations \citep{Marcote2017}.}
    \label{fig:121102_fit}
\end{figure}

\begin{figure}
    \centering
    \includegraphics[width=\linewidth]{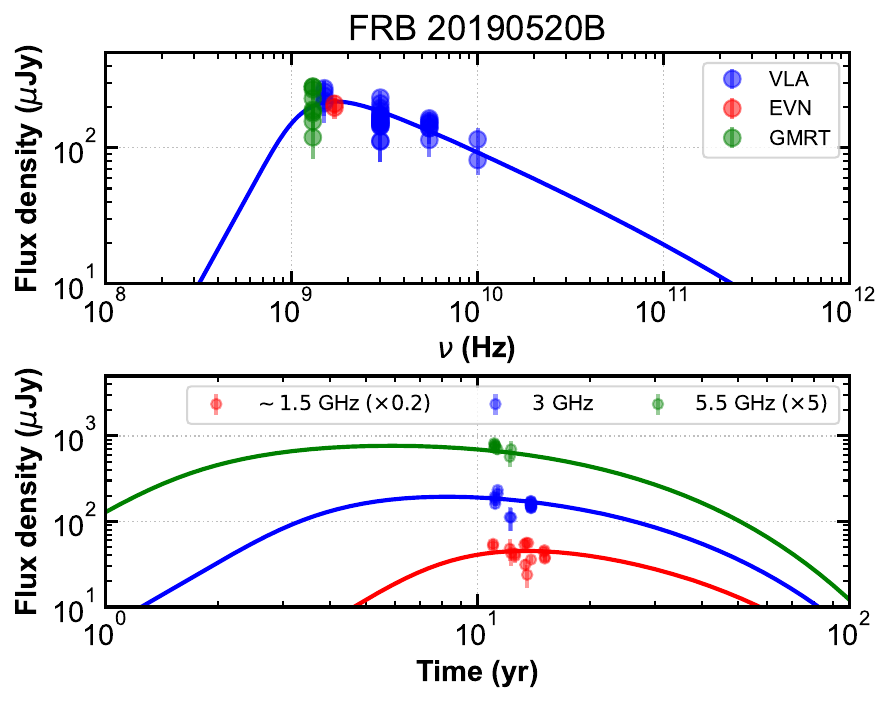}
    \caption{The fitting results of SEDs (top panel) and light curves (bottom panel) of PRS associated with FRB 20190520B. The observation data for the different observation epochs are listed in Table \ref{tab:prs_obs}. The observation data of PRS associated with FRB 20190520B at $\nu\sim 1.5,3,5.5$ GHz are taken from Table 1 in \cite{Balasubramanian2025}, and the light curves with $\nu\sim 1.5,3,5.5$ GHz are shown in red, blue and green lines in the bottom panel. A young magnetar with $t_{\mathrm{age}}=11$ yr, $B_{\mathrm{int}}=3\times 10^{16}$ G and $t_0=50$ yr can explain the observed PRS for FRB 20190520B. Other parameters are listed in Table \ref{tab:results}. The age obtained by fitting the PRS flux is consistent with the conclusions of previous work, which are roughly a dozen years old, whether through the analysis of the PRS flux \citep{Margalit2018,Zhao2021,Bhattacharya2025} and DM/RM variations \citep{Zhao2021a,Wang2025}. The physical size of the emission region is $R_{\mathrm{m}}\sim \chi v_{\mathrm{n}}t=0.11$ pc for FRB 20190520B, which satisfies the given constraint of $<9$ pc of VLBI observations \citep{Bhandari2023}.}
    \label{fig:190520_fit}
\end{figure}

FRB 20121102A and FRB 20190520B are the first and second FRBs associated with compact PRSs \citep{Chatterjee2017,Niu2022}, and also the two sources that have been most well-studied in observations. PRS associated with FRB 20121102A exhibits a flat spectrum in the frequency range of 400 MHz to 22 GHz \citep{Chatterjee2017,Marcote2017,Resmi2021,Plavin2022,Rhodes2023,Bhardwaj2025,Yang2024}. From the seven-year temporal monitoring of the PRS associated with FRB 20121102A at $L$-band, no significant luminosity variation is found \citep{Bhardwaj2025}. PRS associated with FRB 20190520B has been detected from 1.3 to 10.0 GHz \citep{Niu2022,Zhang2023,Bhandari2023,Balasubramanian2025}. The PRS luminosity is decreasing over the period of $\sim 4$ yr at frequencies of $1.5-5.5$ GHz \citep{Zhang2023,Balasubramanian2025}. Our fitting results of SEDs (top panel) and light curves (bottom panel) of PRS associated with FRB 20121102A and FRB 20190520B are given in Figure \ref{fig:121102_fit} and \ref{fig:190520_fit}, respectively. The observation data for the different observation epochs are listed in Table \ref{tab:prs_obs}. The temporal $L$-band (1.3-1.7 GHz) observations of PRS associated with FRB 20121102A are taken from Table 2 in \cite{Bhardwaj2025}, and the light curve with $\nu=1.5$ GHz is shown in red lines in the bottom panel of Figure \ref{fig:121102_fit}. The observation data of PRS associated with FRB 20190520B at $\nu\sim 1.5,3,5.5$ GHz are taken from Table 1 in \cite{Balasubramanian2025}. The light curves with $\nu\sim 1.5,3,5.5$ GHz are shown in red, blue and green lines in the bottom panel of Figure \ref{fig:190520_fit}. 

We find a young magnetar with $t_{\mathrm{age}}=15$ yr and $B_{\mathrm{int}}=2.7\times 10^{16}$ G ($t_{\mathrm{age}}=11$ yr and $B_{\mathrm{int}}=3\times 10^{16}$ G) can explain the observed PRS for FRB 20121102A (FRB 20190520B). The age obtained by fitting the PRS flux is consistent with the conclusions of previous work, which are roughly a dozen years old, whether through the analysis of the PRS flux \citep{Margalit2018,Zhao2021, Bhattacharya2025} and DM/RM variations \citep{Zhao2021a,Wang2025}. The magnetization parameter $\sigma_{\mathrm{B}}=0.1$ is used \citep{Margalit2018}. The temperatures of the injected electrons are $k_{\mathrm{B}}T_{\mathrm{e}}\simeq 1.3$ GeV and $k_{\mathrm{B}}T_{\mathrm{e}}\simeq 0.92$ GeV for FRB 20121102A and FRB 20190520B, respectively. The physical size of the emission region is $R_{\mathrm{m}}\sim \chi v_{\mathrm{n}}t=0.15(0.11)$ pc for FRB 20121102A (FRB 20190520B), which satisfies the given constraint of $<0.7$ pc ($<9$ pc) of VLBI observations \citep{Marcote2017,Bhandari2023}.

\subsubsection{FRB 20201124A and FRB 20240114A}
\begin{figure}
    \centering
    \includegraphics[width=\linewidth]{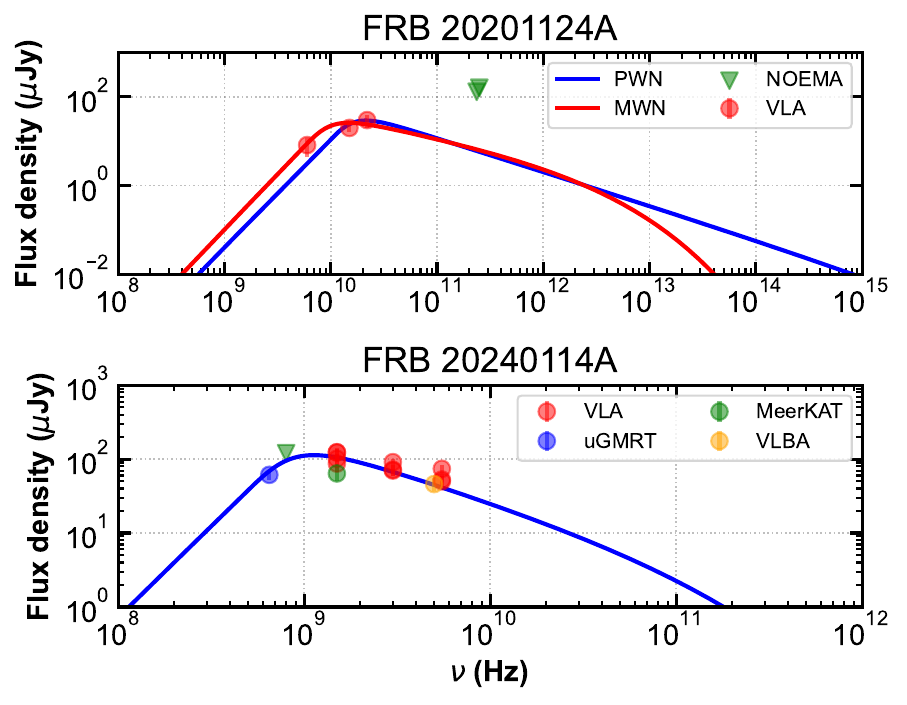}
    \caption{SEDs of PRS associated with FRB 20201124A (top panel) and FRB 20240114A (bottom panel). The observation data for the different observation epochs are listed in Table \ref{tab:prs_obs}. \textbf{Top panel:} The rotation-powered nebula (blue line) and the magnetic-powered nebula (red line) can explain the observed PRS for FRB 20201124A. For PWN(or MWN) models, the PRS is powered by a magnetar with $B_{\mathrm{d}}=10^{14}$ G, $P_{\mathrm{i}}=50$ ms (or $B_{\mathrm{int}}=1\times 10^{16}$ G, $t_0=10^2$ yr ). We found that the magnetar producing FRB 20201124A is the youngest of all PRSs, with an age of 7.2 years. \textbf{Bottom panel:} For FRB 20240114A, the observed PRS can be powered by a magnetar with $t_{\mathrm{age}}=12$ yr, $B_{\mathrm{int}}=2.3\times 10^{16}$ G and $t_0=10^3$ yr. The size of the emission region is $R_\mathrm{m}\sim \chi v_{\mathrm{n}}t\sim 0.06$ pc, which is broadly consistent with the radius $\sim 0.03$ pc constrained from the SSA \citep{Zhang2025}.}
    \label{fig:201124_fit}
\end{figure}

Recently, PRS has been detected in association with both FRB 20201124A \citep{Bruni2024} and FRB 20240114A \citep{Bruni2025}. The PRS associated with FRB 20201124A exhibits an inverted spectrum (with a positive spectral index $F_\nu \propto \nu^\alpha, \alpha=1$) at higher frequencies of 6-22 GHz \citep{Bruni2024}. The PRS associated with FRB 20240114A is similar to that of FRB 20121102A and FRB 20190520B, exhibiting a flat spectrum \citep{Bruni2025,Bhusare2025,Zhang2025}. The spectral break around $\sim 1.6$ GHz has been detected for the PRS associated with FRB 20240114A, which may be caused by the SSA \citep{Zhang2025}. From the SSA frequency, the size of the emission region is $\sim 0.03$ pc \citep{Zhang2025}. Our fitting results of SEDs of PRS associated with FRB 20201124A (top panel) and FRB 20240114A (bottom panel) are given in Figure \ref{fig:201124_fit}. 

The luminosity of the PRS associated with FRB 20240114A is much lower than that of FRB 20121102A and FRB 20190520B, a rapidly rotating magnetar ($B_{\mathrm{d}}=10^{14}$ G, $P_{\mathrm{i}}=50$ ms) is sufficient to drive the PWN with the observed flux density (blue line). Alternatively, the PRS associated with FRB 20240114A could be driven by a magnetar with a strong magnetic field ($B_{\mathrm{int}}=1\times 10^{16}$ G, see red line). Within the framework of the expanding nebula model, a higher radiation peak frequency implies a younger source age (see Figure \ref{fig:rot} and \ref{fig:mag}). We found that the magnetar producing FRB 20201124A is the youngest of all PRSs, with an age of 7.2 years. The size of the emission region is $R_{\mathrm{m}}\sim \chi v_{\mathrm{n}}t=0.002$ pc with the nebula expanding velocity $v_{\mathrm{n}}=3\times 10^{8}$ cm s$^{-1}$ and $\chi=0.1$.

Radio signals may be absorbed by dense ejecta for young SNRs. For an oxygen-dominated homologously expanding ejecta, the SNR becomes optically thin to radio emission at an age of 7.2 years ($\tau_{\rm ff,ej}(t=7.2 ~\rm yr)<1$), which requires \citep{Metzger2017}
\begin{equation}
f_{\mathrm{ion}} M_{\mathrm{ej}} \lesssim 0.46 M_{\odot} \bar{g}_{\mathrm{ff}}^{-1 / 2} \nu_9 T_{\rm ej,4}^{3 / 4} v_{\mathrm{ej}, 9}^{5 / 2},
\end{equation}
where $f_{\rm ion}$ is the ionization fraction of the ejecta and $T_{\rm ej}$ is the ejecta temperature. This condition is readily satisfied for core-collapse SNe (e.g., $M_{\mathrm{ej}} \lesssim 4.6 M_{\odot}$ for $f_{\mathrm{ion}}=0.1$). A detailed modeling of the optical depth and the contributed DM of the SNR is presented in \cite{Wang2025}.

Based solely on the observed spectrum of PRS associated with FRB 20201124A, we cannot distinguish between the PWN and MWN (see Figure \ref{fig:201124_fit}). We present here predictions for the higher frequency spectrum and light curves, which can be tested by subsequent observations. For the rotation-powered model, the injected electrons obey a broken power-law distribution with a broken Lorentz factor $\gamma_{\mathrm{b}}=5\times 10^4$. For magnetic-powered models, injected electrons are assumed to be a relativistic Maxwellian distribution with the temperature $k_{\mathrm{B}}T_{\mathrm{e}}\simeq 0.18$ GeV. The emission from thermal electrons results in an exponential cut-off at frequencies above $\nu \sim 100$ GHz. In the case of rotation-powered nebula models, the flux of the PRS will decay rapidly over time. While in the case of magnetic-powered nebula models, the flux of the PRS can remain stable for a considerable period. 

For FRB 20240114A, the observed PRS can be powered by a magnetar with $t_{\mathrm{age}}=12$ yr, $B_{\mathrm{int}}=2.3\times 10^{16}$ G and $t_0=10^3$ yr. The temperature of injected electrons is $k_{\mathrm{B}}T_{\mathrm{e}}\simeq 0.25$ GeV. The size of the emission region is $R_\mathrm{m}\sim \chi v_{\mathrm{n}}t\sim 0.06$ pc, which is broadly consistent with the radius $\sim 0.03$ pc constrained from the SSA \citep{Zhang2025}.

\subsection{Multiwavelength Radiation from the nebula}
IC scattering of synchrotron photons in the nebula may produce multi-wavelength radiation. The distance for the located FRB–PRS systems is $\sim400-1200$ Mpc \citep{Chatterjee2017,Niu2022,Bruni2024,Bruni2025}. At such distances, detecting high-energy radiation is extremely challenging because of the low expected fluxes. In this section, we calculate the SSC spectra based on the best-fitting synchrotron radio spectra of observed PRSs (see Section \ref{sec:prs_fitting}), and estimate the maximum distances at which the corresponding high-energy emission could be detected with current instruments.

\begin{figure}
    \centering
    \includegraphics[width=\linewidth]{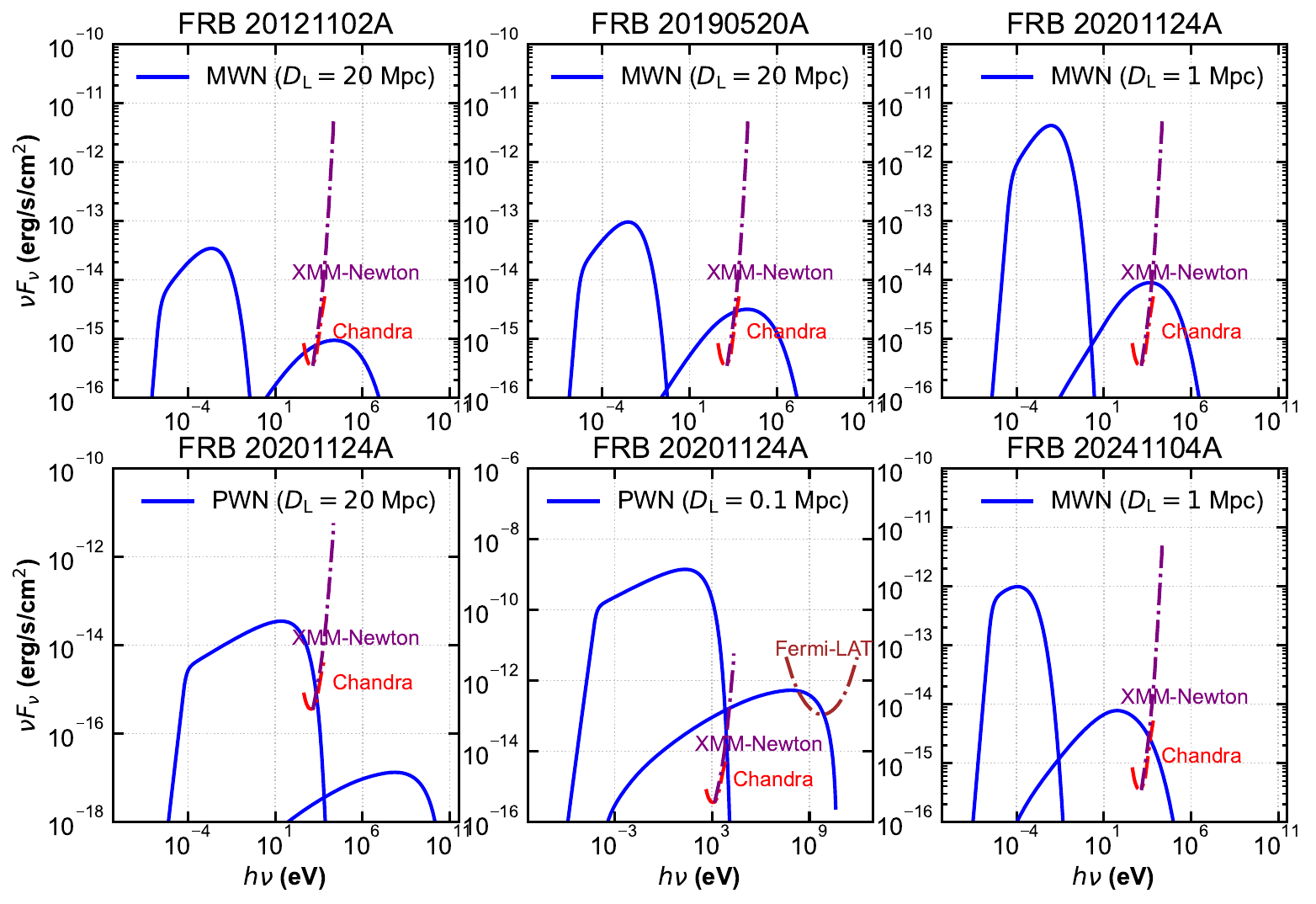}
    \caption{The synchrotron radiation and SSC scattering spectrum of observed PRSs. The synchrotron radio spectra are taken from our best-fitting results (see Section \ref{sec:prs_fitting}). For magnetic-powered models, the X-ray counterparts can be detected only if the PRSs are located at distances of $D_{\mathrm{L}}=1-20$ Mpc. For rotation-powered models, the synchrotron radiation produces broadband emissions from radio to X-ray. For FRB 20201124A, the X-ray emissions through synchrotron radiation from the nebula can be detected if it is located at a distance of $D_{\mathrm{L}}\sim 20$ Mpc. The SSC process mainly produces the emission at the $\sim$GeV band, which can be detected at distances of $D_{\mathrm{L}}=0.1$ Mpc.}
    \label{fig:PRS_SSC}
\end{figure}

The broadband spectrum of synchrotron radiation and SSC scattering is shown in Figure \ref{fig:PRS_SSC}. For magnetic-powered models, the characteristic energies of SSC photons are
\begin{equation}
    \epsilon_{\mathrm{SSC}}\simeq 4 \langle\gamma_{\mathrm{e}}\rangle^2 \epsilon_{\mathrm{SYN}}\approx 2~ \mathrm{keV} \left(\frac{\langle\gamma_{\mathrm{e}}\rangle}{500}\right)^2\left(\frac{\nu_{\mathrm{SYN}}}{5~\mathrm{GHz}}\right).
\end{equation}

The X-ray counterparts of PRSs can be detected only if the PRSs are located at distances of $D_{\mathrm{L}}=1-20$ Mpc. For rotation-powered models, the synchrotron radiation produce broadband emissions from radio to X-ray. For FRB 20201124A, the X-ray emissions through synchrotron radiation from the nebula can be detected if it is located at distances of $D_{\mathrm{L}}\sim 20$ Mpc. The SSC process mainly produces the emission at the $\sim$GeV band, which can be detected at distances of $D_{\mathrm{L}}=0.1$ Mpc.

\section{DM and RM from the nebula}\label{sec:dmrm}
The local environments of FRBs can be probed through variations in DM and RM. The complex and diverse evolution of DM and RM has been observed in active repeating FRBs. For FRB 20121102A, the DM increases before 2020
\citep{Hessels2019,Oostrum2020,Li2021}, and then declines \citep{WangP2025,Snelders2025}. The RM of FRB 20121102A has been found to have extreme values RM$\sim 10^5$ rad m$^{-2}$ \citep{Michilli2018},  and a 70\% decrease from its initial observation \citep{Hilmarsson2021,WangP2025,Zhang2026}. FRB 20190520B has been found embedded in a dense environment, characterized by an extreme DM exceeding the contributions from both the intergalactic medium and the Milky Way, amounting to approximately $900 \mathrm{~pc} \mathrm{~cm}^{-3}$ \citep{Niu2022}. The DM of FRB 20190520B declines with a rate of $\sim12$ pc cm$^{-3}$ yr$^{-1}$ \citep{Niu2022}. The rapid RM variations or reversals have been observed for some active FRBs, such as FRB 20201124A \citep{Xu2022}, FRB 20190520B \citep{Anna-Thomas2023}, FRB 20190417A \citep{Feng2025,Moroianu2026} and FRB 20220529 \citep{Li2026}.

One-zone models struggle to explain complex DM and RM variations. Recently, \cite{Wang2025} proposed a unified model, the magnetar-massive-star binary embedded in a SNR, which can explain most DM and RM observations. As a supplement to the unified model, we will discuss the DM and RM contributed by relativistic electrons in the nebula. In this work, we only focus on the relativistic electrons given by the solution of Equation (\ref{eq:Ngamma}). 


The main components of pulsar winds are relativistic electron-positron pairs, which do not contribute to RM. The magnetar outflows consist primarily of ion-electron plasma, whose contributions to DM and RM are expressed as follows: 
\begin{equation}
\begin{gathered}
\mathrm{DM}_{\mathrm{MWN}}=R_{\mathrm{m}} \cdot \int \frac{n_{e, \gamma}}{\gamma_\mathrm{e}} d \gamma_\mathrm{e}, \\
\mathrm{RM}_{\mathrm{MWN}}=\frac{e^3}{2 \pi m_e^2 c^4} R_{\mathrm{m}} \langle B_{\mathrm{m}}\rangle \cdot \int \frac{n_{e, \gamma}}{\gamma_\mathrm{e}^2} d \gamma_\mathrm{e},
\end{gathered}
\end{equation}
where $\langle B_{\mathrm{m}}\rangle$ is the averaged magnetic field of the magnetized region. 

\begin{figure}
    \centering
    \includegraphics[width=\linewidth]{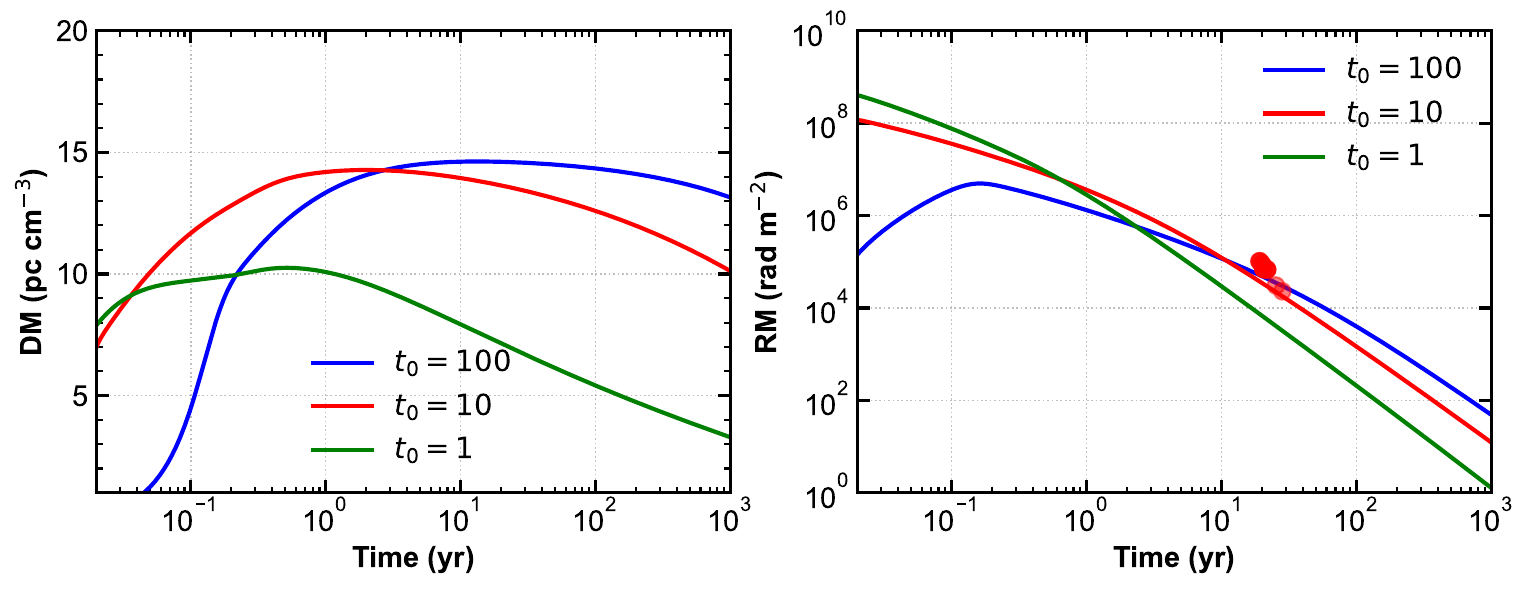}
    \caption{The evolution of DM (left panel) and RM (right panel) from the nebula of FRB 20121102A. Blue, red and green lines represent the cases of $t_0=100,10,1$ yr. Other parameters are the same as Figure \ref{fig:121102_fit}. The DM contributed by the nebula is only a few dozen pc cm$^{-3}$. Moreover, significant decay occurs only when $t \gg t_0$, which contradicts the high activity of FRB 20121102A. The extreme and rapidly decreasing RM of FRB 20121102A may be caused by the magnetized nebula. The red circles are the observation values taken from \cite{Michilli2018,Hilmarsson2021,WangP2025}.}
    \label{fig:121102_rm}
\end{figure}

The evolution of DM (left panel) and RM (right panel) from the nebula of FRB 20121102A is shown in Figure \ref{fig:121102_rm}. Blue, red and green lines represent the cases of $t_0=100,10,1$ yr. Other parameters are the same as Figure \ref{fig:121102_fit}. The DM contributed by the nebula is only a few dozen pc cm$^{-3}$. Moreover, significant decay occurs only when $t \gg t_0$, which contradicts the high activity of FRB 20121102A. The observed variations in DM are unlikely to be caused by MWN, but are more likely dominated by the binary system or SNR \citep{Wang2025}.

The extreme and rapidly decreasing RM of FRB 20121102A may be caused by the magnetized nebula. The red circles are the observation values taken from \cite{Michilli2018,Hilmarsson2021,WangP2025}. Our results can broadly reproduce the measured RM values and their evolution trend. However, precise fitting requires further consideration of the contributions from other environments, as \cite{Wang2025} pointed out. It is challenging to account for FRBs with RM reversals using the expanding nebula model. We do not address this issue in this paper, as an explanation has been previously provided in \cite{Wang2025}.

\section{Conclusions}\label{sec:con} 
The persistent radio counterparts have been detected for some active repeating FRBs. The long-term evolution of the DM and RM of these FRBs is diverse. One-zone models encounter challenges in accounting for intricate DM and RM variations. Recently, \cite{Wang2025} proposed a unified model, the magnetar-massive-star binary embedded in a SNR, which can explain PRS, DM and RM observations. In this work, we investigate how the observed spectra and light curves of PRSs can be explained within the framework of a binary system. In addition, we have also provided the multi-wavelength counterparts from the wind interaction between the magnetar and the massive star. Our conclusions are as follows:
\begin{itemize}
\item For a young magnetar to drive FRBs and bright PRSs, the magnetar wind is usually stronger than the stellar wind. In the direction perpendicular to the orbital plane, the magnetar wind will propagate freely, leading to the formation of the MWN during the interaction between the magnetar's outflow and the SN ejecta. In the orbital plane, the bow shock will form at the standoff distance determined by the pressure balance of the magnetar wind and the stellar wind.
\item Considering the radiation of the bow shock is governed by the IC process, the emission is mainly at high energy band. Based on the estimation of the emission efficiency, we find that the bow shock radiation can only account for some faint PRSs with luminosity $\nu L_{\nu}\lesssim 10^{36}$ erg s$^{-1}$.
\item The bright PRS may originate from the synchrotron radiation from the MWN at a distance much greater than the orbit separation. The MWN may be powered by the rotational energy or the internal magnetic energy. The spindown timescale of a magnetar is very short, resulting in the faster decay of the luminosity. The rapid spin-down limits the luminosity of PRS to only reach above $10^{29}$ erg/s/Hz when they are extremely young (e.g., $t_{\mathrm{age}} < 10 $ yr, see Figure \ref{fig:rot}(1b)). However, the SNR is expected to be optically thick for radio signals at an age $\lesssim 10$ yr due to the free-free absorption of SN ejecta \citep{Metzger2017,Zhao2021}. 
\item Compared to rotational energy, the internal magnetic energy of magnetars has a longer decay timescale, which allows the luminosity of MWNs to remain stable over a period $1-100$ yr. The observed spectra of PRSs associated with FRB 20121102A, FRB 20190520B and FRB 20240114A can be explained by the synchrotron radiation from the magnetic-powered nebula with internal magnetic field $\sim 10^{16}$ G. The ages of FRB 20121102A, FRB 20190520B and FRB 20240114A are a $10-20$ yr, which is consistent with the study of DM and RM variations \citep{Wang2025}.
\item The PRS luminosity is stable for FRB 20121102A during the seven-year monitoring \citep{Bhardwaj2025}, while it is decreasing over the period of $\sim 4$ yr for FRB 20190520B \citep{Zhang2023,Balasubramanian2025}.
For FRB 20121102A, the magnetic field decay timescale is 100 yr. When the age is much smaller than the decay timescale, the luminosity can remain stable. FRB 20190520B has a shorter decay timescale ($t_0=50$ yr), the luminosity is decreasing when $t_{\mathrm{age}}\sim t_0$.
\item Both PWN and MWN can explain the observed spectrum of the PRS associated with FRB 20201124A. We present here predictions for the higher frequency spectrum and light curves, which subsequent observations can test. For magnetic-powered models, the emission from thermal electrons results in an exponential cut-off at frequencies above $\nu \sim 100$ GHz. In the case of rotation-powered nebula models, the flux of the PRS will decay rapidly over time. While in the case of magnetic-powered nebula models, the flux of the PRS can remain stable for a considerable period.
\item In magnetic-powered models, SSC scattering in the nebula may produce keV X-ray emission, whereas in rotation-powered models it may produce GeV $\gamma$-ray emission. The X-ray and $\gamma$-ray emissions can be detected by current instruments at distances of $\lesssim 20$ Mpc and $\lesssim 0.1$ Mpc, respectively.
\item Magnetars associated with active FRBs are younger and more active. Therefore, if they are in massive binary systems, the wind interaction will produce much stronger multi-wavelength radiation than that of Galactic gamma-ray binaries. The synchrotron keV radiation can be detected by Chandra and XMM-Newton at the distance of $\lesssim 50$ Mpc. The SSC and EIC scattering radiation is mainly at the GeV band, which can be detected by Fermi-LAT at the distance of $\lesssim 1$ Mpc. The VHE TeV $\gamma$-ray photons can be detected by \textit{CTA} and \textit{LHAASO} at the distance of $\lesssim 0.1$ Mpc. 
\end{itemize}

\begin{acknowledgments}
We thank the anonymous referee for helpful comments. Z.Y.Z. thanks the helpful discussions with Yu-Jia Wei and Yue Wu. This work was supported by the National Natural Science Foundation of China (grant Nos. 12494575, 12393812, 12273009, 12447115 and 12503050). Q.W. is supported by the China Postdoctoral Science Foundation (CPSF) under Grant Number GZB20240308, 2025T180875 and 2025M773199.
\end{acknowledgments}

\bibliographystyle{aasjournalv7}
\bibliography{bibtex.bib}



\clearpage
\appendix

\section{List of Symbols}
The symbols used in this paper are listed in Table \ref{tab:symbols}.
\begin{table}
\centering
\caption{Summary of the main symbols used in the physical models.}
\label{tab:symbols}
\begin{tabular}{llll}
\hline
Category & Symbol & Meaning & Model \\
\hline
Central engine 
& $j=\{\mathrm{sd},B\}$ 
& Engine index: spin-down 
& All models \\

& &or magnetic-field decay & \\

& $\mathcal{E}_{\mathrm{inj}}=\int \dot{E}_j dt$ & Injected energy & PWN/MWN\\

Rotation-powered  
& $P_{\rm i}$, $\Omega_{\rm i}$ 
& Initial spin period and angular frequency 
& Rotation-powered model \\

engine & $B_{\rm d}$ 
& Dipolar magnetic field 
& Rotation-powered model \\

& $E_{\rm rot}$ 
& Initial rotational energy 
& Rotation-powered model \\

& $L_{\rm sd}$, $t_{\rm sd}$ 
& Spin-down luminosity and spin-down timescale 
& Rotation-powered model \\

Magnetic-powered  
& $B_{\rm int}$ 
& Initial internal magnetic field 
& Magnetic-powered model \\

engine & $E_{\rm B,i}$ 
& Initial internal magnetic energy 
& Magnetic-powered model \\

& $L_{\rm B}$, $L_{\rm B,0}$ 
& Magnetic-field-decay luminosity and initial value 
& Magnetic-powered model \\

& $t_\mathrm{d}$, $\alpha_\mathrm{B}$ 
& Magnetic-field-decay timescale and decay index 
& Magnetic-powered model \\

Microphysics 
& $\sigma_j$ 
& Magnetization parameter of the injected outflow 
& All models\\

& $\gamma_{\min}$, $\gamma_b$, $\gamma_{\max}$ 
& Minimum, break, and maximum  
& Rotation-powered model \\

& &Lorentz factors of injected electrons &\\

& $p_1$, $p_2$ 
& Low- and high-energy indices  
& Rotation-powered model \\

& & of the broken power-law electron spectrum & \\


& $\Theta_{\mathrm{e}}=k_{\mathrm{B}} T_{\mathrm{e}}/m_{\mathrm{e}} c^2$
& Dimensionless electron temperature 
& Magnetic-powered model \\

& $\langle \gamma_e \rangle$
& Mean electron Lorentz factor
& All models \\

& $\langle \epsilon_j \rangle$ 
& Mean injected particle energy 
& All models \\

Dynamics 
& $t_{\rm age}$ 
& Source age at the discovery epoch 
& Source-specific fits \\

& $M_{\rm ej}$, $\mathcal{E}_{\rm SN}$ 
& SN ejecta mass and explosion energy 
& PWN/MWN \\

& $R_{\rm ej}$, $v_{\rm ej}$ 
& Ejecta radius and velocity 
& PWN/MWN \\

& $R_{\rm n}$, $v_{\rm n}$ 
& Nebular radius and expansion velocity 
& PWN/MWN \\

& $R_{\rm m}=\chi R_{\rm n}$ 
& Size of the magnetized emitting region 
& PWN/MWN \\

Binary and  
& $M_{\rm m}$, $M_\star$ 
& Magnetar mass and companion-star mass 
& Binary model \\

wind interaction & $R_\star$, $T_\star$
& Companion radius, effective temperature 
& Binary model \\

& $L_\star$ 
& Companion luminosity 
& Binary model \\

& $\dot{M}$, $v_{\rm w}$ 
& Stellar-wind mass-loss rate and velocity 
& Binary model \\

& $P_{\rm orb}$, $d$ 
& Orbital period and orbital separation 
& Binary model \\

& $\eta$ 
& Momentum-flux ratio between 
& Binary model \\

& & magnetar wind and stellar wind 
& \\

& $r_\mathrm{s}$, $R_\mathrm{s}$ 
& Shock distances from 
& Binary model \\
&  
& the magnetar and companion star 
& \\

Radiation  
& $L_\nu^{\rm syn}$, $L_\nu^{\rm SSC}$ 
& Synchrotron and SSC spectral luminosity 
& All models \\

and opacity &  $L_\nu^{\rm EIC}$ 
& External IC spectral luminosities 
& Binary model \\

& $Y$ 
& Compton parameter 
& Binary model \\

& $\tau_{\rm SSA}$ 
& Synchrotron self-absorption optical depth 
& All models \\

& $\tau_{\rm ff,w}$ 
& Free-free optical depth of the stellar wind 
& Binary model \\

& $\tau_{\gamma\gamma}$ 
& Pair-production optical depth 
& Binary model \\
\hline
\end{tabular}
\end{table}

\clearpage

\section{The Stellar Model}\label{sec:star}
The stellar properties mainly depend on the stellar type. The stellar luminosity and radius can be given by the mass-luminosity relation and mass–radius relation \citep{Salaris2005,Iben2012}
\begin{equation}
L_{\star,\odot} \simeq\left\{\begin{array}{ll}
0.23 M_{\star,\odot}^{2.3} & M_\star \leq 0.43 M_{\odot} \\
M_{\star,\odot}^4 & 0.43 M_{\odot}<M_\star \leq 2 M_{\odot} \\
1.4 M_{\star,\odot}^{3.5} & 2 M_{\odot}<M_\star \leq 55 M_{\odot} \\
32000 M_{\star,\odot} & M_\star>55 M_{\odot}
\end{array},\right.
\end{equation}
\begin{equation}
R_{\star,\odot}\simeq\left\{\begin{array}{ll}
M_{\star,\odot}^{0.8} & M_\star \leq 1 M_{\odot} \\
M_{\star,\odot}^{0.57} & M_\star>1 M_{\odot}
\end{array} ,\right.
\end{equation}
where $L_{\star,\odot}=L_\star/L_\odot$, $M_{\star,\odot}=M_\star/M_\odot$ and $R_{\star,\odot}=R_\star/R_\odot$
are the luminosity, mass and radius of the star in units of solar luminosity, mass and radius. The mass-loss rate of massive stars can be described as a function of the surface effective temperature \citep{Vink2000}: for O-type stars, the mass-loss rate is
\begin{equation}
\begin{aligned}
&\begin{aligned}
\log \left(\frac{\dot{M}}{M_\odot \mathrm{yr}^{-1}}\right)= & -6.697  +2.194 \log \left(\frac{L_\star}{10^5~L_\odot}\right) -1.313 \log \left(\frac{M_\star}{30~M_\odot}\right) -1.226\log \left(\frac{v_{\infty} / v_{\text{esc}}}{2.0}\right) \\
&+0.933 \log \left(\frac{T_\star}{4\times10^4 ~\mathrm{K}}\right) -10.92\left[\log \left(\frac{T_\star}{4\times10^4 ~\mathrm{K}}\right)\right]^2
\end{aligned}\\
&\text { for } 27500<T_\star \leq 50000 \mathrm{~K},
\end{aligned}
\end{equation}
and for B-type stars is 
\begin{equation}
\begin{aligned}
&\begin{aligned}
\log \left(\frac{\dot{M}}{M_\odot \mathrm{yr}^{-1}}\right)= & -6.688 +2.210\log \left(\frac{L_\star}{10^5~L_\odot}\right) -1.339\log \left(\frac{M_\star}{30~M_\odot}\right) \\
& -1.601 \log \left(\frac{v_{\infty} / v_{\mathrm{esc}}}{2.0}\right) +1.07 \log \left(\frac{T_\star}{2\times10^4 ~\mathrm{K}}\right)
\end{aligned}\\
&\text { for } 12500<T_{\text {eff }} \leq 22500 \mathrm{~K}.
\end{aligned}
\end{equation}
$v_{\text {esc}}=\left(2 G M_\star / R_\star\right)^{1 / 2}$ is the escaping velocity of the star. The stellar wind velocity at the distance $r$ is
\begin{equation}
v_{\mathrm{w}}=v_{\infty}\left(1-\frac{R_\star}{r}\right)^{1 / 2},
\end{equation}
where $v_{\infty}$ is the stellar wind velocity at $r\rightarrow \infty$, and wind velocity is $v_{\mathrm{w}} \sim v_{\infty}$ for the distance far away from the star ($r\gg R_\star$). The velosity $v_{\infty} / v_{\mathrm{esc}}=2.6$ for O-type stars and $v_{\infty} / v_{\mathrm{esc}}=1.3$ for B-type stars is used \citep{Vink2000}. For the star with $M_{\star}\lesssim 3.2 M_\odot(T_{\star}\lesssim 12500~\mathrm{K})$, solar mass-loss rate $\dot{M}_{\odot} \approx 2 \times 10^{-14} M_{\odot}~\mathrm{yr}^{-1}$ is used \citep{Wood2002}. 

\section{The radiation efficiency}\label{sec:ic}
The luminosity ratio of the IC radiation (including SSC and EIC) to the synchrotron radiation can be described by the Compton $Y$ parameter \citep{Sari2001}
\begin{equation}\label{eq:Y}
Y \equiv \frac{L_{\mathrm{SSC}}+L_{\mathrm{EIC}}}{L_{\mathrm{SYN}}}=\frac{U_{\mathrm{syn}}+U_{\mathrm{star}}}{U_\mathrm{B}},
\end{equation}
where $U_{\mathrm{syn}}$, $U_{\mathrm{star}}$ and $U_{\mathrm{B}}$ are the energy densities of synchrotron radiation, stellar radiation and magnetic field at the location of the bow shock
\begin{equation}
\begin{aligned}
& U_{\mathrm{syn}}=\frac{\epsilon}{1+Y} \frac{\dot{E}}{4 \pi  (1+\sigma)c r_{\mathrm{s}}^2}, \\
& U_{\mathrm{star}}=\frac{L_\star}{4 \pi cR_{\mathrm{s}}^2}, \\            
& U_{\mathrm{B}}=\frac{B_\mathrm{b}^2}{8 \pi},
\end{aligned}
\end{equation}
where $\epsilon=\epsilon_{\mathrm{rad}}\epsilon_{\mathrm{KN}}$ is 
the radiation efficiency. The efficiency $\epsilon_{\mathrm{rad}}=\min \left\{1,\left(\gamma_{\mathrm{m}} / \gamma_{\mathrm{c}}\right)^{(p-2)}\right\}$ is the fraction of the electron energy radiated away \citep{Sari2001}. For the photons with a frequency $\nu>\nu_{\mathrm{KN}}=m_{\mathrm{e}} c^2 /(h \gamma_{\mathrm{e}})$, the IC scattering enters the Klein–Nishina (KN) regime. The IC radiation should be suppressed with efficiency $\epsilon_{\mathrm{KN}}$ \citep{Wang2010} 
\begin{equation}
\epsilon_{\mathrm{KN}}\left(\gamma_{\mathrm{e}}\right)= \begin{cases}\left(\frac{\nu_{\mathrm{m}}}{\nu_{\mathrm{c}}}\right)^{\frac{3-p}{2}}\left[\frac{\nu_{\mathrm{KN}}\left(\gamma_{\mathrm{e}}\right)}{\nu_{\mathrm{m}}}\right]^{4 / 3}, & \nu_{\mathrm{KN}}\left(\gamma_{\mathrm{e}}\right) \leq \nu_{\mathrm{m}} \\ {\left[\frac{\nu_{\mathrm{KN}}\left(\gamma_{\mathrm{e}}\right)}{\nu_{\mathrm{c}}}\right]^{\frac{3-p}{2}},} & \nu_{\mathrm{m}}<\nu_{\mathrm{KN}}\left(\gamma_{\mathrm{e}}\right)<\nu_{\mathrm{c}} \\ 1, & \nu_{\mathrm{c}} \leq \nu_{\mathrm{KN}}\left(\gamma_{\mathrm{e}}\right) ,\end{cases}
\end{equation}
being the fraction of photons below the KN limit for the slow-cooling case ($\gamma_{\mathrm{e,m}}<\gamma_{\mathrm{e,c}}$), and 
\begin{equation}
\epsilon_{\mathrm{KN}}\left(\gamma_{\mathrm{e}}\right)= \begin{cases}\left(\frac{\nu_{\mathrm{c}}}{\nu_{\mathrm{m}}}\right)^{1 / 2}\left[\frac{\nu_{\mathrm{KN}}\left(\gamma_{\mathrm{e}}\right)}{\nu_{\mathrm{c}}}\right]^{4 / 3}, & \nu_{\mathrm{KN}}\left(\gamma_{\mathrm{e}}\right) \leq \nu_{\mathrm{c}} \\ {\left[\frac{\nu_{\mathrm{KN}}\left(\gamma_{\mathrm{e}}\right)}{\nu_{\mathrm{m}}}\right]^{1 / 2},} & \nu_{\mathrm{c}}<\nu_{\mathrm{KN}}\left(\gamma_{\mathrm{e}}\right)<\nu_{\mathrm{m}} \\ 1, & \nu_{\mathrm{m}} \leq \nu_{\mathrm{KN}}\left(\gamma_{\mathrm{e}}\right),\end{cases}
\end{equation}
for the fast-cooling case ($\gamma_{\mathrm{e,c}}<\gamma_{\mathrm{e,m}}$). The frequencies $\nu_{\mathrm{c}}$ and $\nu_{\mathrm{m}}$ are the typical synchrotron radiation frequencies $\nu(\gamma_{\mathrm{e}})=3 \gamma_{\mathrm{e}}^2 e B /(4 \pi m_{\mathrm{e}} c)$ for the Lorentz factors $\gamma_{\mathrm{e}}=\gamma_{\mathrm{e,c}}$ and $\gamma_{\mathrm{e}}=\gamma_{\mathrm{e,m}}$, respectively. The stellar luminosity is $L_\star=4 \pi R_\star^2 \sigma_{\mathrm{SB}} T_\star^4$, which is mainly determined by the stellar type. 

Solving equation (\ref{eq:Y}), the Compton parameter $Y$ is
\begin{equation}
    Y = \frac{1}{2}\left[(\tilde{b}-1)+\sqrt{(\tilde{b}-1)^2+4(\tilde{a}+\tilde{b})}\right],
\end{equation}
where 
\begin{equation}
        \tilde{a} = \frac{2 \epsilon}{\sigma}\approx 200 \sigma_{-2}^{-1},
\end{equation}
and 
\begin{equation}
\begin{aligned}
\tilde{b} &= Y_{\mathrm{EIC}}=\frac{2 L_\star \eta(1+\sigma)}{\sigma \dot{E} }=\frac{2 L_\star (1+\sigma)}{\dot{M} v_\mathrm{w} c \sigma}\\
&\approx 350 L_{\star,38}\sigma_{-2}^{-1}\left(\frac{\dot{M}}{10^{-7} M_\odot \mathrm{yr}^{-1}}\right)^{-1} \left(\frac{v_{\mathrm{w}}}{3\times 10^8~\mathrm{cm~s^{-1}}}\right)^{-1}.
\end{aligned}
\end{equation}
The EIC parameter $Y_{\mathrm{EIC}}$ depends only on the star's properties. If the EIC scattering can be neglected (e.g., for low-mass companions), we can obtain results similar to those of \cite{Sari2001}.

\section{Radiation Mechanism}\label{sec:rad}
The spectral power of synchrotron radiation at frequency $\nu$ is \citep{Rybicki1979}
\begin{equation}\label{eq:Psyn}
P_\nu^{\mathrm{SYN}}(\gamma_{\mathrm{e}})=\frac{\sqrt{3} e^3 B}{m_{\mathrm{e}} c^2} F\left(\frac{\nu}{\nu_{\mathrm{ch}}}\right), \quad F(x)=x \int_x^{+\infty} K_{5 / 3}(k) dk,
\end{equation}
where $\nu_{\mathrm{ch}}=3\gamma_{\mathrm{e}}^2 e B / (4 \pi m_{\mathrm{e}} c)$ is the characteristic frequency and $K_{5 / 3}(k)$ is modified Bessel function of order 5/3. The emissivity and absorption coefficients of synchrotron radiation for the electron spectrum $n_{\mathrm{e}, \gamma}$ are
\begin{equation}
\begin{gathered}
j_\nu^{\mathrm{SYN}}=\frac{1}{4 \pi}\int n_{\mathrm{e}, \gamma} P_\nu^{\mathrm{SYN}}(\gamma_{\mathrm{e}}) d \gamma_{\mathrm{e}}, \\
\alpha_\nu^{\mathrm{SYN}}=-\frac{1}{8 \pi m_{\mathrm{e}} \nu^2}\int \gamma_{\mathrm{e}}^2 P_\nu^{\mathrm{SYN}}(\gamma_{\mathrm{e}})\frac{\partial}{\partial \gamma_{\mathrm{e}}}\left(\frac{n_{e, \gamma}}{\gamma_{\mathrm{e}}^2}\right) d \gamma_{\mathrm{e}}.
\end{gathered}
\end{equation}
The SSA optical depth is $\tau_{\mathrm{SSA}}=\alpha_\nu R$, where $R$ is the radius of the radiation region.

The SSC radiation power at frequency $\nu$ is given by \citep{Blumenthal1970}
\begin{equation}\label{eq:Pssc}
\begin{aligned}
P_\nu^{\mathrm{SSC}}(\gamma_{\mathrm{e}})&=3 \sigma_{\mathrm{T}} \int_{\nu_{\mathrm{s}, \min }}^{\infty}  \frac{\nu f_{\nu_{\mathrm{s}}}^{\mathrm{SYN}}}{4 \gamma_{\mathrm{e}}^2 \nu_{\mathrm{s}}^2} g(x, y)\mathrm{d} \nu_{\mathrm{s}}, \\
g(x, y)&=2 y \ln y+(1+2 y)(1-y)+\frac{x^2 y^2}{2(1+x y)}(1-y),
\end{aligned}
\end{equation}
where $\nu_{\mathrm{s}}$ is the frequency of seed synchrotron radiation photons, $\nu_{\mathrm{s}, \min }=\nu m_{\mathrm{e}} c^2 /\left[4 \gamma_{\mathrm{e}}\left(\gamma_{\mathrm{e}} m_{\mathrm{e}} c^2-h \nu\right)\right]$, $x=4 \gamma_{\mathrm{e}} h \nu_{\mathrm{s}} /\left(m_{\mathrm{e}} c^2\right)$ and $y=h \nu /\left[x \left(\gamma_{\mathrm{e}} m_{\mathrm{e}} c^2-h\nu\right)\right]$.

The EIC radiation power at frequency $\nu$ is given by \citep{Aharonian1981}
\begin{equation}\label{eq:Peic}
\begin{aligned}
& P_\nu^{\mathrm{EIC}}\left(\gamma_{\mathrm{e}}, \mu\right)=3 \sigma_{\mathrm{T}} \int_{\nu_{\mathrm{s}, \min }}^{\infty}  \frac{\nu f_{\nu_{\mathrm{s}}}^{\mathrm{STAR}}}{4 \gamma_{\mathrm{e}}^2 \nu_{\mathrm{s}}^2} h\left(\xi, b_\theta\right) \mathrm{d} \nu_{\mathrm{s}}\\
& h\left(\xi, b_\theta\right)=1+\frac{\xi^2}{2(1-\xi)}-\frac{2 \xi}{b_\theta(1-\xi)}+\frac{2 \xi^2}{b_\theta^2(1-\xi)^2},
\end{aligned}
\end{equation}
where $\mu=\cos \theta$ is the scattering angle, $\xi=h \nu /\left(\gamma_{\mathrm{e}} m_{\mathrm{e}} c^2\right)$ and $b_\theta=2\left(1-\mu\right) \gamma_{\mathrm{e}} h \nu_{\mathrm{s}} /\left(m_{\mathrm{e}} c^2\right)$. The flux density of the synchrotron radiation (the star) photons is
\begin{equation}
f_{\nu_{\mathrm{s}}}^{\mathrm{SYN}}=\frac{L_{\nu_{\mathrm{s}}}^{\mathrm{SYN}}}{4\pi R^2}, \quad f_{\nu_{\mathrm{s}}}^{\mathrm{STAR}}=\pi B_{\nu_{\mathrm{s}}}(T_\star)\left(R_\star / R\right)^2,       
\end{equation}
where $L_{\nu_{\mathrm{s}}}^{\mathrm{SYN}}$ is the synchrotron radiation luminosity and $B_{\nu_{\mathrm{s}}}\left(T_\star\right)=2 h \nu_{\mathrm{s}}^3 / c^2\left[\exp \left(h \nu_{\mathrm{s}}/ k T_\star\right)-1\right]$ is the brightness of the companion. The high-energy photons produced by the IC process could be absorbed by the photon field via $\gamma \gamma$ annihilation \citep{Gould1967}. The cross section of $\gamma \gamma$ annihilation effects is
\begin{equation}
\begin{aligned}
    \sigma_{\gamma \gamma}&=\frac{3 \sigma_\mathrm{T}}{16}\left(1-\beta^2\right)\left[\left(3-\beta^4\right) \ln \frac{1+\beta}{1-\beta}-2 \beta\left(2-\beta^2\right)\right],\\
    \beta&=\sqrt{1-\frac{2 m_e^2 c^4}{h\nu h\nu_{\mathrm{s}}(1-\mu)}}.
\end{aligned}
\end{equation}
The absorption via $\gamma \gamma$ annihilation from the bow shock is mainly from the stellar photon field, and the optical depth is
\begin{equation}
\tau_{\gamma \gamma}=\frac{R}{2} \int_{-1}^1(1-\mu) d \mu \int \sigma_{\gamma \gamma} n_{\gamma}\left(\nu\right) d \nu,
\end{equation}
where $n_{\gamma}$ is the photon density from the companion.

The total synchrotron and IC radiation specific luminosity is 
\begin{equation}
L_\nu^{\mathrm{tot}}=L_\nu^{\mathrm{SYN}} \frac{1-e^{-\tau_{\mathrm{SSA}}}}{\tau_{\mathrm{SSA}}}+(L_\nu^{\mathrm{SSC}}+L_\nu^{\mathrm{EIC}}) e^{-\tau_{\gamma \gamma}}.
\end{equation}
If the radiation region is uniform and homogeneous, the luminosity for mechanism A can be expressed as
\begin{equation}
L_\nu^{\mathrm{A}}=4 \pi \int_V j_\nu^{\mathrm{A}} d V= \int N_{\mathrm{e},\gamma} P_\nu^{\mathrm{A}}(\gamma_{\mathrm{e}})d \gamma_{\mathrm{e}}, 
\end{equation}
where $N_{\mathrm{e},\gamma}=n_{\mathrm{e},\gamma}V$ is the electron number distribution. The spectral power $P_\nu^{\mathrm{A}}$ for synchrotron, SSC and EIC are given in equations (\ref{eq:Psyn}), (\ref{eq:Pssc}) and (\ref{eq:Peic}). 

\clearpage

\begin{table}[]
    \centering
    \begin{tabular}{ccccc}
    \hline
     Source    &  Observation epoch & Central Frequency& Telescope & References\\
     &  (YYYY) & (GHz) &  & \\
    \hline
     FRB 20121102A    & 2016 & 1.63-22 & VLA & \cite{Chatterjee2017}\\
        & 2016 & 1.67 & EVN & \cite{Chatterjee2017}\\
        & 2016 & 1.55-4.98 & VLBA& \cite{Chatterjee2017}\\
        & 2016 & 1.7-5.0 & EVN& \cite{Marcote2017}\\
        & 2017 & 0.4-1.39 & GMRT& \cite{Resmi2021}\\
        & 2017 & 1.7-4.8 & EVN& \cite{Plavin2022}\\
        & 2019,2022 & 1.28 & MeerKAT& \cite{Rhodes2023}\\
        & 2022-2023 & 1.26 & GMRT& \cite{Bhardwaj2025}\\
        & 2023 & 1.5 & VLA& \cite{Yang2024}\\   
     \hline
     FRB 20190520B & 2020 & 1.5-5.5 & VLA & \cite{Niu2022}\\
      & 2020-2021 & 1.5-10.0 & VLA & \cite{Zhang2023}\\
     & 2022 & 1.7 & EVN & \cite{Bhandari2023}\\
    & 2022-2024 & 1.3-3.0 & GMRT & \cite{Balasubramanian2025}\\
    & 2023-2024 & 1.5-3.0 & VLA & \cite{Balasubramanian2025}\\
    \hline
   FRB 20201124A & 2021-2022 & 6.0-22.0 & VLA & \cite{Bruni2024}\\
    \hline
   FRB 20240114A & 2024 & 5.0 & VLBA & \cite{Bruni2025}\\
  & 2024 & 0.65 & GMRT & \cite{Bhusare2025}\\
  & 2024 & 1.5-5.5 & VLA & \cite{Zhang2025}\\
  & 2024 & 1.5 & MeerKAT & \cite{Zhang2025}\\
  \hline
    \end{tabular}
    \caption{Observations of PRSs.}
    \label{tab:prs_obs}
\end{table}

\begin{table}[]
    \centering
    \begin{tabular}{ccccc}
    \hline
    Source &  Age &  MWN models  &  PWN models  &  Dynamics parameters\\
          &  $t_{\mathrm{age}}$ &  ($B_{\mathrm{int}},t_0,\sigma_{\mathrm{B}},k_{\mathrm{B}}T_{\mathrm{e}}$) &  ($B_{\mathrm{d}},P_{\mathrm{i}},\sigma_{\mathrm{B}},\gamma_{\mathrm{b}}$) & ($v_{\mathrm{n}},\chi$)\\
          &  yr &  ($10^{16}$ G, yr, 1, GeV) &  ($10^{14}$ G, ms, 1, 1) &  ($10^8$ cm s$^{-1}$, 1)\\
    \hline
    FRB 20121102A  & 15 & (2.7, 100, 0.1, 1.3) & &   (10, 1)\\
    FRB 20190520B   & 11 & (3, 50, 0.1, 0.92) & &   (10, 1)\\
    FRB 20201124A   & 7.2 &  & (1, 50, 0.05, $5\times 10^4$)&   (3, 0.1)\\
    FRB 20201124A & 7.2 & (1, 100, 0.1, 0.18) & &   (3, 0.1)\\
    FRB 20240114A  & 12 & (2.3, 10$^3$, 0.1, 0.25) & &  (5, 1)\\
    \hline
    \end{tabular}
    \caption{Fitting parameters for the MWN/PWN models}
    \label{tab:results}
\end{table}

\end{document}